\documentclass[a4paper,11pt]{article}
\usepackage{jheppub_mod}
\usepackage{hyperref}
\usepackage{amsfonts,amssymb,amsmath}
\usepackage{color}
\usepackage{pdfsync}
\usepackage{accents}
\usepackage{diagbox}
\usepackage{yfonts}   
\usepackage{lscape}
\usepackage{rotating,graphicx}
\usepackage[nottoc,notlot,notlof]{tocbibind}

\usepackage{soul}
\usepackage{cancel}
\usepackage{verbatim}
\usepackage{appendix}
\usepackage[T1]{fontenc} 
\newcommand{\AAA}{\mathcal A}

\newcommand{\CCC}{\mathcal C}
\newcommand{\EEE}{\mathcal E}
\newcommand{\HHH}{\mathcal H}
\newcommand{\KKK}{\mathcal K}
\newcommand{\MMM}{\mathcal M}

\newcommand{\TTT}{\mathcal T}
\newcommand{\ZZZ}{\mathcal Z}
\newcommand{\C}{\mathbb C}

\newcommand\F{\mathbb{F}}
\renewcommand{\H}{\mathbb H}
\newcommand{\N}{\mathbb N}

\newcommand{\R}{\mathbb R}
\newcommand{\Z}{\mathbb Z}
\newcommand{\Zt}{\mathbb{Z}_2}

\newcommand{\bps}{{\rm BPS}}
\newcommand{\NS}{{\rm NS}}
\newcommand{\Ra}{{\rm R}}
\newcommand{\Rat}{\widetilde{{\rm R}}}

\newcommand{\Tr}{{\rm Tr}}
\newcommand\wh{\widehat}
\DeclareFontFamily{U}{mathx}{\hyphenchar\font45}
\DeclareFontShape{U}{mathx}{m}{n}{
      <5> <6> <7> <8> <9> <10>
      <10.95> <12> <14.4> <17.28> <20.74> <24.88>
      mathx10
      }{}
\DeclareSymbolFont{mathx}{U}{mathx}{m}{n}
\DeclareFontSubstitution{U}{mathx}{m}{n}
\DeclareMathAccent{\widecheck}{0}{mathx}{"71}
\DeclareMathAccent{\wideparen}{0}{mathx}{"75}

\newcommand{\hf}{\frac{1}{2}}
\newcommand{\qt}{\frac{1}{4}}

\newcommand{\kop}{(\chi^1_+)}
\newcommand{\ktp}{(\chi^2_+)}
\newcommand{\kom}{(\chi^1_-)}
\newcommand{\ktm}{(\chi^2_-)}
\newcommand{\kbop}{(\qu{\chi}^1_+)}
\newcommand{\kbtp}{(\qu{\chi}^2_+)}

\newcommand{\aop}{(a^1_+)}
\newcommand{\atp}{(a^2_+)}
\newcommand{\aom}{(a^1_-)}
\newcommand{\atm}{(a^2_-)}
\newcommand{\geo}{SU(2)_{{\rm geom}}}
\newcommand{\lgeo}{\overline{SU(2)}_{{\rm geom}}}

\newcommand{\ds}{\displaystyle}
\newcommand{\qu}{\overline}

\newcommand{\nop[1]}{\mathop{:\!#1\!\!:}}
\newcommand{\be}{\begin{equation}}
\newcommand{\ee}{\end{equation}}
\newcommand{\ba}{\begin{eqnarray}}
\newcommand{\ea}{\end{eqnarray}}
\definecolor{purple}{rgb}{0.6,0,0.6}
\definecolor{bl}{rgb}{0.2,0.1,0.9}
\definecolor{tur}{rgb}{0.1,0.9,0.9}
\definecolor{gr}{rgb}{0.2,0.7,0.2}
\definecolor{pink}{rgb}{1,0,0.8}

\newcounter{save}
\newcounter{rmk}

\numberwithin{equation}{section}
\makeatletter
\newcommand*\rel@kern[1]{\kern#1\dimexpr\macc@kerna}
\newcommand*\widebar[1]{%
  \begingroup
  \def\mathaccent##1##2{%
    \rel@kern{0.8}%
    \overline{\rel@kern{-0.8}\macc@nucleus\rel@kern{0.2}}%
    \rel@kern{-0.2}%
  }%
  \macc@depth\@ne
  \let\math@bgroup\@empty \let\math@egroup\macc@set@skewchar
  \mathsurround\z@ \frozen@everymath{\mathgroup\macc@group\relax}%
  \macc@set@skewchar\relax
  \let\mathaccentV\macc@nested@a
  \macc@nested@a\relax111{#1}%
  \endgroup
}

\title{\boldmath \mbox{}\hfill {\small DCPT-19/21}\\[30pt]
$SU(2)$ channels the cancellation of K3 BPS states}
\vspace{ -1cm}

\author[a,1]{A. Taormina \note{Corresponding author.}}
\author[b]{and K. Wendland}

\affiliation[a] {Centre for Particle Theory, Department of Mathematical Sciences,\\
Durham University, Stockton Road, Durham, DH1 3LE, U.K.}
\affiliation[b] {Mathematics Institute,  Albert-Ludwigs-Universit\"at Freiburg,\\
Ernst-Zermelo-Str. 1, D-79104 Freiburg, Germany}

\emailAdd{anne.taormina@durham.ac.uk}
\emailAdd{katrin.wendland@math.uni-freiburg.de}

\abstract{
The conformal field theoretic elliptic genus, an invariant for $N=(2,2)$ superconformal field theories, 
counts the BPS states in any such theory with signs, according to their bosonic or fermionic nature.  
For K3 theories, this invariant is the source of the Mathieu Moonshine phenomenon. There, the  
net number of $\qt$-BPS states is positive for any conformal dimension above the massless threshold, 
but it may arise after cancellation of the contributions of an equal number of bosonic and fermionic BPS 
states present in  non-generic theories, as is the case for the class of $\mathbb{Z}_2$-orbifolds of 
toroidal SCFTs. Nevertheless, the space
$\widehat\HHH$ of all BPS states that are generic to such orbifold theories provides a 
convenient framework to construct a particular
generic space of states of K3 theories. We find a 
natural action 
of the group $SU(2)$ on a subspace of  $\widehat\HHH$ which is 
compatible with the cancellations of contributions from the corresponding non-generic states. 
In fact, we propose that this action channels those cancellations. As a by-product, we find 
a new subspace of the generic space of states in  $\widehat\HHH$.}

\begin{document}
 \maketitle
 \flushbottom

\section{Introduction}
The BPS spectrum of quantum field theories with extended supersymmetry has long
been recognised to yield crucial and in fact useful information. Indeed, it is the key to the 
construction of invariants which both allow extrapolations from weak  to strong coupling 
and relate abstractly defined theories to geometry.  K3 theories provide a rich family of examples. 
They are superconformal field theories with $N=(2,2)$ worldsheet supersymmetry at central charges 
$(c, \qu{c})=(6,6)$, with spacetime supersymmetry and integer eigenvalues of the operators $J_0$ 
and $\qu J_0$ --  the zero modes of the two $\mathfrak{u}(1)$ currents generating an affine  
subalgebra of the left and right $N=2$ superconformal algebras -- and without
holomorphic BPS states at weight $1\over2$. These requirements imply that every 
K3 theory has $N=(4,4)$ worldsheet supersymmetry, and that its conformal field theoretic 
elliptic genus $\EEE(\tau,z)$ equals the complex elliptic genus of K3 surfaces \cite{eguchi1989superconformal}.
The moduli space $\MMM_{K3}$ of K3 theories has dimension $80$ and possesses at least
one connected component whose structure is well understood \cite{aspinwall1994string,nahm2001hiker} and which
contains all $\Z_2$-orbifold conformal field theories that are obtained from toroidal CFTs \cite{nahm2001hiker}. 
This component is expected to parametrize all non linear sigma models on 
K3 surfaces. These, in turn,
provide a rich environment in which to explore non perturbative effects. 

More recently, the sporadic group Mathieu 24 ($M_{24}$) has made an intriguing  appearance in the 
conformal field theoretic elliptic genus of K3 theories. First  came a numerological observation 
by \cite{eguchi2011notes} quickly followed by further evidence from the calculation of twining genera 
\cite{cheng2010k3, gaberdiel2010amathieu, gaberdiel2010bmathieu,eguchi2011note}, culminating in a proof that the numerology is truly a 
signature of $M_{24}$ \cite{gannon2016much}. The prospect of having this sporadic group acting on a non linear 
sigma model on K3 was swiftly discarded in \cite{gaberdiel4315symmetries}, prompting a finer inspection of the BPS spectrum 
of such theories, since $\EEE(\tau,z)$ counts $\hf$- and $\qt$-BPS states with signs, according to their 
bosonic or fermionic nature. In particular, this index encodes a net number of massive $\qt$-BPS ground states, 
all of the same statistics.  It is an invariant throughout $\MMM_{K3}$
and therefore one is at liberty to explore its properties from any vantage point in that moduli space. 
It is known that for generic theories the net number is actually the total number of these
$\qt$-BPS states 
\cite{song2017chiral,wendland2019hodge}, while non-generic theories, to which the $\mathbb{Z}_2$-orbifold CFTs belong, typically 
possess a number of bosonic and fermionic $\qt$-BPS states  whose contributions cancel out
in $\EEE(\tau,z)$. These states are therefore not encoded in $\EEE(\tau,z)$ and  are not 
expected to be organised in non-trivial representations of $M_{24}$. Understanding the nature of these 
 ``excess'' states in order to better grasp the role of $M_{24}$ in relation to the elliptic genus of K3 is the 
 object of the present work. 

Although generic theories do not have such excess states and would therefore 
appear to be less complicated, no such theories are known explicitly. Hence they 
offer little scope for elucidating the $M_{24}$ action on the BPS states counted by $\EEE(\tau,z)$ at present. 
In contrast, $\mathbb{Z}_2$-orbifold CFTs provide an interesting laboratory, not the least because they all 
share the same spectrum of generic BPS states; in other words, the fine details of their underlying complex
2-tori are of no consequence for our analysis of this class of K3 theories. Moreover,
these theories enjoy a wealth of beautiful mathematical structures, which are only beginning to show. Indeed, in this work
we use a global $SU(2)$ action on a subspace of generic BPS states
which we hope will provide a useful tool in the study of K3 theories beyond
the known examples, and certainly beyond Mathieu Moonshine. 

A few years ago, and guided by our symmetry surfing programme \cite{taormina2010symmetries, taormina2013overarching, taormina2015symmetry}, we showed, 
in the framework of $\mathbb{Z}_2$-orbifold CFTs, that a maximal subgroup of $M_{24}$ called the 
\emph {octad group} $G$ acts naturally on a space of $\qt$-BPS
states at level  one\footnote{In the Ramond sector, states at level $n \in \N$ have  conformal 
dimensions  $(h, \qu{h})=(\frac{1}{4}+n, \qt)$.} whose dimension agrees with the 
massive $\qt$-BPS contribution to 
the elliptic genus at level  one  \cite{taormina2015twist}.
Significantly, the octad group is not a subgroup of $M_{23}$ and is
the  overarching group of all geometric symmetry groups $G_i$ of different $\Z_2$-orbifold
CFTs on K3 \cite{taormina2015symmetry}. These geometric symmetries are rooted in the construction
of Kummer surfaces, obtained by minimally resolving the $16$ singularities of the standard
$\Zt$-quotient of a complex $2$-torus $T_{\Lambda}=\C^2/\Lambda$ with $\Lambda$ a rank $4$ 
lattice.
In the symmetry surfing programme, the ability to surf  the $\Zt$-orbifold subvariety 
of $\MMM_{K3}$ relies crucially on the existence, in the rank $16$ Kummer lattice, of a ``diagonal'' direction invariant 
under the action of  all geometric symmetry groups $G_i$. On the $\mathbb{Z}_2$-orbifold CFT 
side, this is echoed by the existence of a ``diagonal'' exactly marginal state  $T^{{\rm diag}}$ 
built on a twisted ground state $|\alpha_{{\rm diag}}\rangle$
whose orthogonal complement in the $16$-dimensional space of twisted ground 
states is a $15$-dimensional subspace $\AAA$. We denote by $H^\perp$ the space of
all massive ground states in the Fock space over $\AAA$. $H^\perp$ is graded and we write
$H^\perp=\bigoplus\limits_{n=1}^\infty H^\perp_n$ where  $H^\perp_n$ is the space of 
level $n$ states in $H^\perp$, $n\in\N$. At level $1$, it so happens that the net number of 
massive $\qt$-BPS states 
counted by the elliptic genus corresponds to states solely built on twisted ground states 
in $\AAA$. Moreover, all the massive $\qt$-BPS states built on $|\alpha_{{\rm diag}}\rangle$
cancel in the counting against all massive $\qt$-BPS states from the untwisted sector. 
This property helped us 
pinning down the octad group action on the remaining massive $\qt$-BPS states at level 1,
using Margolin's construction 
of a $45$-dimensional representation of $M_{24}$ as a guide \cite{margolin1993geometry}, but the situation is not typical 
at higher level. Yet the octad group continues to act on $H^\perp$, as elegantly 
demonstrated by Gaberdiel, Keller and Paul \cite{gaberdiel2017mathieu}. Moreover, 
they argue that $G$  also acts on the 
space $H^{{\rm rest}}$ consisting of some
$\qt$-BPS states built on the twisted ground state  $|\alpha_{{\rm diag}}\rangle$
and some $\qt$-BPS states from the untwisted sector 
such that the graded dimension of $H^{\rm rest}\oplus H^\perp$ agrees precisely with the massive 
ground state contributions to   
$\EEE(\tau,z)$. Since the action of $G$ arises from symmetry surfing, this beautifully 
supports  the symmetry surfing programme.

Let us describe the $\qt$-BPS states in a given K3 theory in more detail. Here and in the following, by this abbreviation
we mean those BPS states which saturate the BPS bound for half of the antiholomorphic $N=4$ worldsheet
supersymmetries and which are massive with respect to the holomorphic $N=4$ superconformal algebra. By 
$\widehat{H}_n$, $n \in \N \setminus\{0\}$,
we denote the space of \emph{all} $\qt$-BPS \emph{ground} states 
at level $n$ in the given K3 theory. The findings of 
\cite{gaberdiel2017mathieu} imply that $\widehat{H}_n\cong H_n^\perp \oplus H_n^{{\rm rest}}\oplus H_n^{+}$, 
where as above, the dimension of $H_n^\perp \oplus H_n^{{\rm rest}}$ agrees with the massive 
ground state contributions at level $n$ to $\EEE(\tau,z)$.
Accordingly, $H_n^+$ accounts for ``excess'' states whose contributions to the conformal field theoretic
elliptic genus cancel out. While $H_n^\perp$ is well under control by the results of \cite{gaberdiel2017mathieu}, the spaces 
$H_n^{{\rm rest}}\oplus H_n^+$ have not yet released all their 
secrets. In particular, for $n>1$
neither $H_n^{{\rm rest}}$ nor $H_n^+$ has been constructed explicitly 
so far in any K3 theory.
 In a very recent article \cite{keller2019lifting}, Keller and Zadeh have deformed  
$\mathbb{Z}_2$-orbifold CFTs on K3 away from the orbifold point using second order 
conformal perturbation methods. They have shown that 
under a deformation by $T^{{\rm diag}}$ all the $\qt$-BPS states in 
$H_1^\perp$ remain $\qt$-BPS, while those in $H_1^+$ cease to satisfy the bound 
once the initial theory is deformed away from the orbifold. 
Deforming away from the orbifold into a different direction lifts the states in a \emph{different}
space $\widetilde H_1^+\cong H_1^+$, whilst $(\widetilde H_1^+)^\perp\subset H_1^+\oplus H_1^\perp$
remains at the BPS bound.
This fits very well in the overall picture of the elliptic genus  providing
information that  remains unchanged regardless of the regime (perturbative or not) one is interested in, 
and regardless of the point in  $\MMM_{K3}$ one considers. It also fits nicely with the symmetry
surfing predictions: depending on the direction of deformation, different symmetry groups remain unbroken.
Hence different subspaces of $\widehat H=\bigoplus\limits_{n=1}^\infty \widehat H_n$ 
remain stable under different deformations, but each  of them is isomorphic,
as a representation of the Virasoro- and $\frak{u}(1)$-current zero modes $L_0,\, \qu L_0,\,
J_0,\, \qu J_0$,
to the massive ground state contributions 
to the generic space of states $\H_0$
 introduced 
in \cite{wendland2019hodge}. This in particular shows that the inclusion
$ H_n^\perp \oplus H_n^{{\rm rest}}\hookrightarrow\widehat H$ of representations
is not uniquely determined in general.

While the structure of $H^\perp$ has been understood by 
the results of \cite{taormina2015twist,gaberdiel2017mathieu}, we now proceed to uncover more structure on the spaces 
$H_n^{{\rm rest}}\oplus H_n^+$, with the hope of finding how  $\Z_2$-orbifold  CFTs earmark 
excess BPS states. 
The results of \cite{wendland2019hodge} show that independently of the deformation, 
if deforming to a generic theory, then the \emph{bosonic} excess
states in $\widehat H_n$ are precisely those states in $\widehat H_n$ that transform in the vacuum representation
of the antiholomorphic $N=4$ superconformal algebra. Under deformation, the $N=4$ representations
built on these states combine with representations with \emph{fermionic} $\qt$-BPS ground states
to form long representations away from the BPS bound. By the above, these excess fermionic ground
states depend on the choice of deformation. To describe them
for the deformation by $T^{\rm diag}$, we construct a ``geometric'' action of the group $SU(2)$, henceforth 
denoted $\geo$, under which the four free Dirac fermions of our theories and 
their superpartners transform as doublets and whose action commutes with that of the $N=4$ 
superconformal algebra.
All $\qt$-BPS states in the graded space 
$H^{{\rm rest}}\oplus H^+=\bigoplus\limits_{n=1}^\infty H^{{\rm rest}}_n\oplus H^+_n$  transform under $\geo$. 
We argue that for deformations in the diagonal direction $T^{{\rm diag}}$, to be lifted to a long 
representation, both the fermionic and bosonic excess $\qt$-BPS states in $H^+$ must transform in  
isomorphic representations of $\geo$. 

Before proceeding to the heart of our matter, we provide the context in which $M_{24}$ emerged 
in K3 theories. This has the double aim of providing a quick overview for the reader who is not familiar 
with the subject, and of introducing some of the definitions and notations that will be used later in the paper.

Let $\tau, z \in \C$,  with $\tau$ in the upper complex halfplane\footnote{By
$\qu\tau, \qu z\in\C$ we denote the complex conjugates of $\tau, z\in\C$, which we occasionally
include as arguments of a non-holomorphic function  of $\tau, z$. This is convenient whenever such
a function allows a power series expansion in $q:=\exp(2\pi i\tau), \qu q:=\exp(-2\pi i\qu \tau), 
y:=\exp(2\pi iz), \qu y:=\exp(-2\pi i\qu z)$, where
$q, \qu q, y, \qu y$ may also be viewed as independent formal variables.}. 
The genus one partition function $\ZZZ^{N=(4,4)}$ of an $N=(4,4)$ superconformal field 
theory  at central charge $c=\qu{c}=6$ is a modular covariant function 
\be \label{partition}
\ZZZ^{N=(4,4)}=\hf\,\left\{\ZZZ^{{\rm NS}}
+\ZZZ^{{\rm R}}+\ZZZ^{{\rm \widetilde{NS}}}
+\ZZZ^{{\rm \widetilde{R}}}\right\}
\ee
with
\ba
\ZZZ^{{\rm S}}
&=&\ZZZ^{{\rm S}}(\tau,z;\qu{\tau}, \qu{z})
:=\Tr_{\H^{\rm S}}\,
(y^{J_0}\,q^{L_0-\qt}\,\qu{y}^{\,\qu J_0}\,
\qu{q}^{\,\qu L_0-\qt}\,),\qquad \qquad \qquad {\rm S}\in \{{\rm NS}, {\rm R}\},\nonumber\\
\ZZZ^{{\rm \widetilde{S}}}
&=&\ZZZ^{{\rm \widetilde{S}}}(\tau,z;\qu{\tau}, \qu{z})
:=\Tr_{\H^{\rm S}}\,((-1)^{J_0-\qu J_0}\,y^{J_0}\,q^{L_0-\qt}\,
\qu{y}^{\,\qu J_0}\,\qu{q}^{\,\qu L_0-\qt}\,) 
\ea
where the traces are taken over the subspaces 
$\H^{{\rm NS}} $ and $\H^{{\rm R}} $ of $\H$, the $(\Zt\times\Zt)$-graded complex vector 
space of all states in the superconformal field theory.  The gradings split $\H$ into the 
Neveu-Schwarz and Ramond sectors\footnote{Referring to both left and right movers,
as in a K3 theory, there exist no $\NS\Ra$ or $\Ra\NS$ states, by definition. 
Note that our description of K3 theories involves the internal CFT only.},
each containing (worldsheet) bosons and fermions, that is
\be
\H=\underbrace{\H_b^{{\rm NS}} \oplus \H_f^{{\rm NS}}}_{\H^{{\NS}}}
\oplus \,\underbrace{\H_b^{{\rm R}}\oplus \H_f^{{\rm R}}}_{\H^{{\rm R}}}.
\ee
Here, as representations of the $N=(4,4)$ superconformal algebras, 
the spaces $\H^{{\NS}}$ and $ \H^{{\rm R}}$ are isomorphically mapped to each other
under spectral flow.
The operators $J_0$ and $\qu J_0$ are the zero modes of the 
$\frak{u}(1)$ currents which, together with the currents $J^\pm(z)$ and 
$\qu J^\pm(\qu{z})$, form two copies of the affine $\frak{su}(2)$  
 subalgebra of the (small) $N=(4,4)$ superconformal algebra. 
The partition function \eqref{partition} may be expressed in terms of sesquilinear  
combinations of $N=4$ unitary irreducible characters at central charges $c=6$ and $\qu{c}=6$. 
These characters  are generating functions for short and long representations and were 
coined `massless' and `massive' respectively in \cite{eguchi1987unitary} to signify that the corresponding 
representations have non-zero or zero Witten index respectively. We recall the expressions for 
these characters in Appendix \ref{subappendix:characters}, where we label the two massless 
characters in the Ramond sector as $\chi_0^{\rm R}(\tau,z)$ and $\chi_\hf^{\rm R}(\tau,z)$ while the Ramond  
massive characters are of the form $q^h\,\widetilde{\chi}^{\rm R}(\tau,z)$ 
with $\widetilde{\chi}^{\rm R}:=\chi_\hf^{\rm R}+2\chi_0^{\rm R}$
and where the conformal weight $h \in \R$ of the highest weight state is bounded below by 
$h >\qt$. The Ramond massless characters, on the other hand, have highest weight states 
whose conformal dimension saturates the
bound $h=\qt$. For K3 theories, the partition function is  of the form
\be
\ZZZ^{N=(4,4)}(\tau,z;\qu{\tau}, \qu{z})
=\hf \sum_{S \in \{\NS, {\rm R}, \widetilde{\NS},\widetilde{\rm R}\}}\,\sum_{a,b}\,n_{ab}\,
\chi_a^{S}(\tau,z)\,\overline{\chi_b^{S}(\tau, z)},\quad n_{ab}\in \N\,\,
\forall a, b,
\ee
with $a,b$ running over massless and massive $N=4$ characters and with the term containing the vacuum of the theory having $n_{00}=1$.
The conformal field theoretic
elliptic genus $\EEE(\tau,z)$ of a K3 theory is defined as the specialisation of 
$\ZZZ^{\widetilde\Ra}$ where the antiholomorphic $\widetilde\Ra$ characters are projected to their Witten index 
value by setting $\qu{z}=0$.  We thus have
\be \label{CFTelliptic}
\EEE(\tau,z):=\Tr_{\H^{{\rm R}}}\,((-1)^{J_0-\qu J_0}\,
y^{J_0}\,q^{L_0-\qt}\qu{q}^{\,\qu L_0-\qt})
\, = \sum_{a,b}\,n_{ab}\,
\chi_a^{\widetilde\Ra}(\tau,z)\,\overline{\chi_b^{\widetilde\Ra}(\tau, 0)}.
\ee
As a consequence of the theory enjoying $N=(2,2)$ worldsheet supersymmetry, 
$\EEE$ is a holomorphic function of $\tau$ and $z$, and it counts (with opposite signs) the 
${\rm RR}$ fermionic 
and bosonic states whose antiholomorphic signature is  the Witten index of the massless 
representation they belong to.  
The first explicit calculation of  this topological invariant was carried out
within the framework of Gepner models and  $\Zt$-orbifold CFTs in 
\cite[(3.8)]{eguchi1989superconformal}, where a spectral-flowed version of the conformal field theoretic
elliptic genus \eqref{CFTelliptic} was used, namely $\Phi(\tau,z):=q^{1/4}y\,\EEE(\tau, z+\frac{\tau +1}{2})$. 
This was in order to make a direct parallel with the work of Witten \cite{witten1987elliptic}. The elliptic genus 
presented in \cite[(5.10)-(5.12)]{eguchi1989superconformal} is the $z=0$ specialisation  of
\be \label{ellipticgenusprime1}
\Phi(\tau,z)
=8\left\{ \frac{\vartheta_2(\tau, z)^2}{\vartheta_4(\tau,0)^2}
-\frac{\vartheta_1(\tau, z)^2}{\vartheta_3(\tau,0)^2}
-\frac{\vartheta_4(\tau, z)^2}{\vartheta_2(\tau,0)^2}\right\}.
\ee
With the help of \eqref{charactersh13} -- \eqref{charactersh42} and \eqref{magic}, 
$\Phi(\tau,z)$  may be expressed as an infinite sum of irreducible $N=4$ Neveu-Schwarz 
characters in the following way,
\be  \label{ellipticgenusphi}
\Phi(\tau,z)
=20\chi_\hf^{{\rm NS}}(\tau,z)
-2\chi_0^{{\rm NS}}(\tau,z)+A(\tau)\,\widetilde{\chi}^{{\rm NS}}(\tau,z),
\ee
where 
$$
A(\tau):=2-8q^{1\over8}\eta(\tau)\sum_{i=2}^{4}h_i(\tau)
$$
and the functions $h_i(\tau)$ are the $\nu=0$ specialisations of the functions $h_i(\tau, \nu)$ given in \eqref{h3} and \eqref{h4h2h1}.
 $A(\tau)$ has Fourier expansion
\be \label{Afunction}
A(\tau)=\sum_{n=1}^{\infty} A_nq^n=2\cdot(45q+231q^2++770q^3+2277q^4+\cdots),
\ee
and $\hf(A(\tau)-2)q^{-{1\over8}}=h^{(2)}(\tau)$, the weakly holomorphic mock modular form on $SL(2,\Z)$ 
presented in \cite[(7.16)]{dabholkar2012quantum}.

A list of the first 8 coefficients $A_n, n\in \{1,..,8\}$, counting $\qt$-BPS ground states,  was given in \cite{ooguri1989superconformal}, 
but the significance of these coefficients  has only been realised since 2010,\footnote{Moonshine, a
clear, unaged whiskey, became legal in the US in 2010. We do not know whether this had
any significance for the discovery of Mathieu Moonshine.} after the observation in \cite{eguchi2011notes} 
that they coincide with dimensions of representations of the sporadic group $M_{24}$. The 
existence of an infinite-dimensional  $M_{24}$ module underlying $A(\tau)$ 
was proven in \cite{gannon2016much}. Yet the role of $M_{24}$ in the context of strings compactified 
on K3 surfaces remains a mystery, and this phenomenon has been  named 
\emph{Mathieu Moonshine}.\\

We have structured the remainder of this work as follows.

In Section \ref{section:2}, we recall the ingredients from $\mathbb{Z}_2$-orbifold CFTs relevant to 
our analysis. Moreover, we present an explicit construction 
of excess BPS states pertaining to 
$H_n^+, n\in \{1,2\}$, including their decomposition into $\geo$ multiplets. 
The information at level $n=1$ was already provided in our  work \cite{taormina2015twist}, but the 
full significance of $\geo$ was not recognised then. The information at level $n=2$ is new 
and requires a careful and detailed analysis of the data encoded in the 
partition function of  $\Z_2$-orbifold CFTs on K3. The states in $H_2^{{\rm rest}}$ are listed 
in Appendix \ref{appendix:states}.

Section \ref{section:3} takes stock of the group theoretic information gleaned in the previous 
section, gives our rationale behind our construction of the $\geo\times\lgeo$ action,
and provides analytic expressions for untwisted and twisted partition functions 
that encode the $\geo$ action. 

A discussion and outlook is given in Section \ref{discussion}.

Appendix \ref{appendix:char} gathers helpful Jacobi theta function identities, as well as 
expressions for the $N=4$ characters at central charge $c=6$ involving Appell functions, 
whose definitions are also presented. This appendix also offers 
explanations for the analytic expressions appearing in Section \ref{section:3}.
\section{$\Zt$-orbifold CFTs}\label{section:2}
Since the elliptic genus $\EEE$  is an invariant on the moduli space $\MMM_{K3}$
of K3 theories, it
encodes properties that all K3 theories share. In particular, apart from 
states in massless representations with respect to the
holomorphic and the antiholomorphic $N=4$ superconformal algebra, which will not be our 
concern here, $\EEE$ counts a net number of $\qt$-BPS states at each integer 
conformal weight strictly above threshold. By the results of \cite{gannon2016much}, the corresponding
contributions to $\EEE$ agree with the graded character of a  space 
\be \label{HBPS}
\HHH^{{\rm BPS}}=\bigoplus_{n=1}^{\infty}\,(H_n\otimes {\cal H}_n^{N=4}),
\ee
where ${\cal H}_n^{N=4}$ is an irreducible massive
$N=4$ representation at conformal weight $n$ for each $n \in\N\setminus\{0\}$, 
and $H_n$ is a finite dimensional representation of $M_{24}$ and $\qu J_0$.
$\HHH^{{\rm BPS}}$ is the subspace of massive states in the generic space of
states $\H_0$ introduced in \cite{wendland2019hodge}.
Each $H_n$ is an invariant  of  K3 theories, although the dimensions of the $\widehat{H}_n$ in the space
\be \label{HhatBPS}
\widehat{\HHH}^{{\rm BPS}}=\bigoplus_{n=1}^{\infty}\,(\widehat{H}_n\otimes {\cal H}^{N=4}_{n})
\ee
of all massive $\qt$-BPS states may vary from one K3 
theory to another. In generic K3 theories, we have $\widehat{H}_n\cong H_n$ for all $n$
\cite{song2017chiral,wendland2019hodge}, while in non-generic theories
${\rm dim}\, \widehat H_n\geq{\rm dim}\,H_n$ for all $n$. When one deforms away from a non-generic 
theory, excess states, whose
contributions to the elliptic genus cancel, are lifted into non-BPS representations off 
threshold.  Although the results of \cite{song2017chiral,wendland2019hodge} imply that 
${\HHH}^{{\rm BPS}}$ has a geometric description in terms of the chiral de Rham cohomology
of K3, this space remains difficult to access. For any deformation of a non-generic theory, 
it is therefore valuable to gain insight on  the subspace of  excess states in
$\widehat{\HHH}^{{\rm BPS}}$ whose contributions to $\EEE$ cancel, and to identify the driver of such cancellations in non-generic 
yet accessible K3 theories.

\subsection{Free fermions and bosons as building blocks} \label{subsection:freefermions}
Our prototype of  non-generic K3
theories is the class of $\Zt$-orbifold superconformal field theories, which we denote by
\be
\CCC=\TTT/\Zt \quad\mbox { with }{\TTT}\mbox{ a
toroidal SCFT at central charges}\,\,  c=\qu c= 6.
\ee   
The construction of $\CCC$ is induced by the standard Kummer construction 
which minimally resolves  the singularities of the $\Zt$-quotient
of a complex $2$-torus $T_\Lambda: = \C^2/\Lambda$, with $\Lambda \subset \C^2$ a rank 4 lattice over $\Z$. 
Unlike generic K3 theories, these provide a framework to test the symmetry surfing idea 
explicitly  and garner further clues for the construction of the putative VOA(s) associated with the 
$M_{24}$ Moonshine module. 
To this effect, we restrict our attention to the symmetry groups $G_i$ induced by geometric 
symmetries - including those stemming from
shifts by half lattice vectors - of the underlying toroidal conformal field theories $\TTT$. 
As was detailed
in \cite{taormina2013overarching,taormina2015twist,taormina2015symmetry}, this is meaningful after the choice of a geometric 
interpretation for the theory $\TTT$ on some torus $T_\Lambda$. 
The symmetry surfing programme also requires a choice of generators for the lattice 
$\Lambda \subset\C^2\cong\R^4$. All in all, these choices induce an
identification ${1\over2}\Lambda/\Lambda\cong\mathbb F_2^4$, such that every geometric
symmetry group $G_i$ acts on the 
twisted ground states $T_{\vec a},\,\vec a\in\F_2^4$, as permutation group
by means of affine linear maps on the space of labels  $\F_2^4$. 

The underlying toroidal CFT ${\cal T}$ possesses 
two holomorphic free Dirac fermions $\chi_+^a(z)$ and their complex conjugates  $\chi_-^a(z)$
with standard OPEs,
\be
\chi_+^a(z)\,\chi_-^b(w) \sim \frac{1}{z-w}\delta^{ab},\quad a,b \in \{1,2\}.
\ee

Their bosonic superpartners $j^a_{\pm}(z)$ are built out of a set of four real holomorphic 
$U(1)$-currents $j^I(z), I\in \{1,2,3,4\}$, whose zero modes generate 
infinitesimal translations on the torus $T_\Lambda$. Here, $j^I(z)$ is the Noether current for 
the translation along the $I^{\rm th}$ coordinate axis in the standard coordinate system that 
$T_\Lambda$ inherits from $\C^2\cong\R^4$ around each point. One has
\be
j^1_\pm(z)=\frac{1}{\sqrt{2}}\,(j^1(z)\pm i j^2(z)),\qquad j^2_\pm(z)=\frac{1}{\sqrt{2}}(j^3(z)\pm i j^4(z)),
\ee
with 
\be \label{bosons}
j^a_+(z)\,j^b_-(w)\sim \frac{1}{(z-w)^2}\,\delta^{ab},\quad a,b \in \{1,2\}.
\ee
Under the action of the $\Zt$-orbifold group, the fields $\chi_\pm^a(z)$ and $j_\pm^a(z)$ flip sign, 
while the $N=4$ SCA is invariant under this orbifold action,
as follows from the following free field representation:
\ba \label{N4}
J^3&=&\frac{1}{2}\{\nop[\chi_+^1\chi_-^1]+\nop[\chi_+^2\chi_-^2]\}
\;=\; \frac{1}{2} J,\qquad 
\phantom{spac}J^\pm=\pm\nop[\chi_\pm^1\chi_\pm^2]\nonumber\\
G^\pm&=&\sqrt{2}\,\{\nop[\chi_\pm^1\,j_\mp^1]+\nop[\chi_\pm^2\,j_\mp^2]\},\qquad \qquad \qquad
G^{\,\prime\pm}=\sqrt{2}\,\{\nop[\chi_\mp^1\,j_\mp^2]-\nop[\chi_\mp^2\,j_\mp^1]\},\\
T&=&\sum_{a=1}^2\nop[j^a_+j^a_-]+\frac{1}{2}\sum_{a=1}^2\{\nop[\partial\chi_+^a\chi_-^a]
+\nop[\partial\chi_-^a\chi_+^a]\}.\nonumber
\ea
The currents $J^\pm$ and $J^3$ generate the $\frak{su}(2)$ affine subalgebra of 
the $N=4$ superconformal algebra, under which 
the Dirac fermions $\chi_\pm^a$  have charges $\pm\frac{1}{2}$, while their bosonic 
superpartners are uncharged, as  is immediate from the form 
of the Cartan subalgebra current $J^3$ in \eqref{N4}. In contrast, the symmetry groups $G_i$ 
also act linearly as subgroups of $SU(2)$ on $\chi_\pm^a$ {\em and} $j_\pm^a$. More precisely, 
$\chi_+^a$ and $j_+^a$  transform as doublets ${\bf 2}$ under
this $SU(2)$, which will be referred to as `geometric' 
$\geo$ while $\chi_-^a$ and $j_-^a$ transform as complex conjugate doublets 
${\bf \qu{2}}$. In other words,  if 
$\chi_+^a$ and $j_+^a$ transform with
\be
M=\begin{pmatrix}
\alpha&\beta\\
- \qu\beta& \qu\alpha
\end{pmatrix}
,\quad \alpha, \beta \in \C,\quad |\alpha|^2+|\beta|^2=1,
\ee
then $\chi_-^a$ and $j_-^a$ transform with $ \qu M$. The action of $\geo$  
commutes with the $N=4$ action, as can be inferred from the $\geo$ invariance of the 
fields in \eqref{N4}.\\

In the antiholomorphic sector of the theory, the two Dirac fermions $\overline{\chi^a_+}(\qu{z})$ and their
superpartners $\overline{j^a_+}(\qu{z})$ transform as doublets under a right-moving
group $\overline{SU(2)}_{{\rm geom}}$
whose action commutes with that of the antiholomorphic $N=4$ superconformal algebra, while they are  singlets under $\geo$.
Their complex conjugates $\overline{\chi^a_-}(\qu{z})$ and $\overline{j^a_-}(\qu{z})$ also transform as doublets
under $\overline{SU(2)}_{{\rm geom}}$ and as singlets under $\geo$.

\subsection{The Neveu-Schwarz partition function}
With an eye to prepare the ground for future work on the VOA(s) expected to
underlie the Mathieu Moonshine module, we choose to work 
in the Neveu-Schwarz sector. The  $\Zt$-orbifold  partition function  in this sector is
given by contributions from the two complex $\NS$ fermions  and their bosonic superpartners 
\eqref{bosons}, both untwisted and twisted by the $\Zt$ action as in \cite{eguchi1989superconformal},
\be
\ZZZ^{\NS} =\ZZZ^{\NS}_{{\rm untwisted}}+\ZZZ^{\NS}_{{\rm twisted}}.
\ee
The dependence on the moduli of the underlying toroidal theory becomes apparent in
$\ZZZ^{\NS}_{{\rm untwisted}}$, which depends on $\Gamma(\Lambda, B)\subset\R^4\oplus\R^4$, 
the signature $(4,4)$ Narain lattice associated with the lattice $\Lambda$ and the $B$-field of the 
underlying toroidal theory. Indeed, we have\footnote{The dependence on $\tau$ will often be 
understood but not explicitly referred to  for  easy reading of formulas.}
\be
\ZZZ^{\NS}_{{\rm untwisted}}(z,\qu{z}) =
\hf \frac{1}{|\eta|^8}\Big\vert\,\frac{\vartheta_3(z)}{\eta}\,\Big\vert^4\,
\big\{1+\sum_{\substack{(p_L;p_R)\in \Gamma(\Lambda, B)\\
(p_L;p_R)\neq (0,0)}}\,q^{\hf p^2_L}\,\qu{q}^{\hf p^2_R}\,\big\}
+8\Big\vert\,\frac{\vartheta_4(z)}{\vartheta_2}\,\Big\vert^4
\ee
and
\be
\ZZZ^{\NS}_{{\rm twisted}}(z,\qu{z}) =
8\Big\vert\,\frac{\vartheta_2(z)}{\vartheta_4}\,\Big\vert^4+8\Big\vert\,\frac{\vartheta_1(z)}{\vartheta_3}\,\Big\vert^4.
\ee
Different tori $T_\Lambda$ lead to different Narain lattices, but the $\qt$-BPS 
states emerging from non-zero momentum or winding are 
non-generic in the class of $\Zt$-orbifold CFTs (see, for example, 
\cite[(3.7)-(3.8)]{wendland2019hodge} for the precise argument). 
The remaining spectrum of states is generic to $\Z_2$-orbifold CFTs on K3, however, 
and is the object of our present analysis.\\

Restricting the graded trace that usually yields the 
Neveu-Schwarz partition function to the states with vanishing winding and momentum
and the twisted sector  thus yields
\be
\ZZZ^{\NS,{\rm generic}}(z,\qu{z})=\hf \frac{1}{|\eta|^8}\Big\vert\,\frac{\vartheta_3(z)}{\eta}\,\Big\vert^4+8\Big\vert\,\frac{\vartheta_4(z)}{\vartheta_2}\,\Big\vert^4+\ZZZ^{\NS}_{{\rm twisted}}(z,\qu{z}).
\ee
The BPS states we are interested in are Neveu-Schwarz states which under the antiholomorphic
$N=4$ superconformal algebra transform like elements of the chiral ring. In other words, we need to
project to $\mbox{ker}\left( 2\qu L_0-\qu J_0\right)$. These states are thus
encoded in the conformal field theoretic elliptic genus of K3 spectral-flowed from 
the $\widetilde{\Ra}$ sector to the $\NS$  sector, that is, in
\be\label{flowedellgendef}
\EEE^{\NS}(\tau, z):=
\Tr_{\H^{{\NS}}}\,((-1)^{\qu J_0}\,
y^{J_0}\,q^{L_0-\qt}\qu{q}^{\,\qu L_0-{\qu J_0\over 2}})
= -q^\qt\,y\,\EEE(\tau, z+\frac{\tau+1}{2}).
\ee
It may also be obtained from the generic Neveu-Schwarz
partition function by inserting $\qu{z}=-\frac{\qu{\tau}+1}{2}$, namely
\ba \label{ellipticNS}
\ZZZ^{\NS,{\rm generic}}(z,\qu{z}
=-\frac{\qu{\tau}+1}{2})
&=&\underbrace{8\frac{\vartheta_4(z)^2}{\vartheta_2^2}\,\qu{q}^{\,-\qt}}_{{\rm untwisted}}
+\underbrace{8\big\{\frac{\vartheta_2(z)^2}{\vartheta_4^2}\,(-\qu{q}^{\,-\qt})
+\frac{\vartheta_1(z)^2}{\vartheta_3^2}\,\qu{q}^{\,-\qt}\big\}}_{{\rm twisted}}\nonumber\\
&=&\EEE^{\NS}(\tau,z)\,\qu{q}^{\,-\qt}.
\ea
Note that our conventions ensure that the vacuum contributes to $\EEE^{\NS}(\tau,z)$
with a positive sign, which we find natural, since it is bosonic\footnote{There is a global sign
difference between equation \eqref{ellipticNS} and \cite[(2.5)]{gaberdiel2017mathieu}, or equivalently, 
\eqref{ellipticgenusprime1}; 
we view this
as a different choice of conventions, due to the fact that in 
\cite{eguchi1989superconformal,gaberdiel2017mathieu}, the elliptic genus
is expressed in the $\NS\widetilde{\rm{R}}$ sector, while we work in the $\NS\NS$ sector.}.

We  now rewrite $\EEE^{\NS}(\tau,z)$ in terms of $N=4$ characters and Appell functions. 
To do so, we  use the following notations introduced in 
\cite[(B.2)--(B.4)]{gaberdiel2017mathieu}: by $U_{\ell=\hf}(z)$ we denote the generating function for BPS states 
that transform as the vacuum under the antiholomorphic $N=4$ superconformal algebra.
To be invariant under the $\Zt$-orbifold action, there must be an even number of modes acting on the vacuum:
\ba \label{Uhf}
U_{\ell=\hf}(z)
&:=&\hf q^{-\qt}\,\Big\{ \prod_{n=1}^\infty\,\frac{(1+q^{n-\hf}y)^2(1+q^{n-\hf}y^{-1})^2}{(1-q^n)^4}+ \prod_{n=1}^\infty\,\frac{(1-q^{n-\hf}y)^2(1-q^{n-\hf}y^{-1})^2}{(1+q^n)^4}\,\Big\}\nonumber\\
&=&\hf \left\{\,\frac{\vartheta_3(z)^2}{\eta^6}+4\frac{\vartheta_4(z)^2}{\vartheta_2^2}\right\}\nonumber\\
&=&\chi_0^{\NS}(z)
+\sum_{n=1}^\infty\,f_n\,q^n\,\widetilde{\chi}^{\NS}(z).
\ea
Similarly,  $U_{\ell=0}(z)$  is the generating function for untwisted BPS states  that transform
as massless matter ground states under the antiholomorphic $N=4$ superconformal algebra.
These states are created from the vacuum by the action of a single mode of weight $\hf$ of an 
antiholomorphic Dirac fermion. Hence for such BPS states to be invariant under the $\Zt$ action, there must be 
an odd number of holomorphic modes acting on the ground state:
\ba \label{U0}
U_{\ell=0}(z)&:=&  
q^{-\qt}\,\Big\{ \prod_{n=1}^\infty\,\frac{(1+q^{n-\hf}y)^2(1+q^{n-\hf}y^{-1})^2}{(1-q^n)^4}- \prod_{n=1}^\infty\,\frac{(1-q^{n-\hf}y)^2(1-q^{n-\hf}y^{-1})^2}{(1+q^n)^4}\,\Big\}\nonumber\\
&=& 
\frac{\vartheta_3(z)^2}{\eta^6}-4\frac{\vartheta_4(z)^2}{\vartheta_2^2}\nonumber\\
&=&4\chi_\hf^{\NS}(z)
+\sum_{n=1}^\infty\,g_n^{\rm inv}\,q^n\,\widetilde{\chi}^{\NS}(z).
\ea
In analogy with our notation \eqref{Afunction}, we introduce 
$$
f(\tau):=\sum_{n=1}^\infty\,f_n\,q^n\quad 
\mbox{ and  }\quad
g^{\rm inv}(\tau):=\sum_{n=1}^\infty\,g_n^{\rm inv}\,q^n, 
$$
and  using \eqref{charactersh42} and \eqref{magic}
we obtain the following analytic expressions for these functions,
\be \label{BtauCtau}
\begin{array}{rcl}
\ds f(\tau)
&=&\ds2h_2(\tau)\,\eta(\tau)\,q^{1\over8}+\frac{q^{1\over8}}{2\eta(\tau)^3}-1,\\[10pt]
\ds  g^{\rm inv}(\tau)
&=& \ds - 4h_2(\tau)\,\eta(\tau)\,q^{1\over8}+\frac{q^{1\over8}}{\eta(\tau)^3}.
\end{array}
\ee
Here, the function $h_2$ is a specialisation of a level one Appell function (see Appendix \ref{subappendix:appell}).

By \eqref{ellipticNS}, the  contributions to
$\EEE^{\NS}(\tau,z)$ from the untwisted sector thus may be written as
\be\label{untwistchar}
8\frac{\vartheta_4(z)^2}{\vartheta_2^2}=2U_{\ell=\hf}(z)-U_{\ell=0}(z)
\stackrel{\eqref{Uhf},\eqref{U0}}{=} 2\chi_0^{\NS}(z)- 4\chi_\hf^{\NS}(z)+(2f-g^{\rm inv})\,\widetilde{\chi}^{\NS}(z).
\ee

Proceeding in a similar fashion in the twisted sector, we use the function $T_{\ell=0}(z)$
introduced in \cite[(B.7)--(B.8)]{gaberdiel2017mathieu} which gives the contributions 
to $\EEE^{\NS}(\tau,z)$ from one  twisted
sector:
 \ba \label{T0}
T_{\ell=0}(z)
&:=&\hf q^{-\qt}\,\Big\{ (y+2+y^{-1})\,q^\hf \,
\prod_{n=1}^\infty\,\frac{(1+q^{n}y)^2(1+q^{n}y^{-1})^2}{(1-q^{n-\hf})^4}\nonumber\\
&&\phantom{spaceseekerspace}+(y-2+y^{-1})
\,q^\hf \, \prod_{n=1}^\infty\,\frac{(1-q^{n}y)^2(1-q^{n}y^{-1})^2}{(1+q^{n-\hf})^4}\,\Big\}\nonumber \\
&=&  \hf\Big\{\frac{\vartheta_2(z)^2}{\vartheta_4^2} -\frac{\vartheta_1(z)^2}{\vartheta_3^2}\Big\}\nonumber\\
&=&\chi_\hf^{\NS}(z)+\sum_{n=1}^\infty\, g^{\rm tw}_n\,q^n\,\widetilde{\chi}^{\NS}(z).
\ea
We introduce the function
\be \label{Dtau}
g^{\rm tw} (\tau)
:=\sum_{n=1}^\infty\,g^{\rm tw}_n\,q^n
\stackrel{\eqref{Tellzero}}{=}-\hf \,(h_3(\tau)+h_4(\tau))\,\eta(\tau)\,q^{1\over8},
\ee
whose Fourier modes, alongside those of  $f(\tau)$ and 
$ g^{\rm inv}(\tau)$, provide crucial data for our analysis. 
By \eqref{ellipticNS} the contributions to $\EEE^{\NS}(\tau,z)$ from the twisted sector thus read
\ba\label{twistchar}
8\Big\{-\frac{\vartheta_2(z)^2}{\vartheta_4^2}+\frac{\vartheta_1(z)^2}{\vartheta_3^2}\Big\}
\stackrel{\eqref{T0}}{=} -16\,T_{\ell=0}(z)
\stackrel{\eqref{T0}}{=} -16 \chi_\hf^{\NS}(z)-16 g^{\rm tw} (\tau) \widetilde{\chi}^{\NS}(z),
 \ea
where the factor $16$ accounts for the number of linearly
independent ground states in the twisted sector of the theory. 
Indeed, the twisted ground states are localised at the 16 singular points of the 
quotient $T_\Lambda/\Zt$. 
Altogether, \eqref{untwistchar} and \eqref{twistchar} yield a decomposition of the 
conformal field theoretic elliptic genus according to
\be \label{ellipticgenusNS}
\EEE^{\NS}(\tau, z) 
= 2 \chi_0^{\NS}(\tau, z) -20 \chi_\hf^{\NS}(\tau, z)
+ \left(2f(\tau) - g^{\rm inv}(\tau) - 16 g^{\rm tw}(\tau)\right) \widetilde{\chi}^{\NS}(\tau, z)
\ee
and thus, by comparison with \eqref{ellipticgenusphi},
\be\label{Adeco}
A(\tau) = -2f(\tau) + g^{\rm inv}(\tau) + 16 g^{\rm tw}(\tau)\,=\sum_{n=1}^\infty A_n q^n.
\ee
With the decomposition \eqref{ellipticgenusNS} of the $\NS$-elliptic genus $\EEE^{\NS}$
in hand, we will from now on continue to work in the Neveu-Schwarz sector, focussing on massive
primary states that contribute to $\EEE^{\NS}$. Note that in this sector, the level $n$ accounted
for by $A_n$ agrees with the conformal weight.
%
\subsection{Decomposition of the space $\widehat{\HHH}^{\bps}$ of 
massive $\qt$-BPS states} \label{subsection:decomposition}
As already mentioned in the introduction to this section, in non-generic K3 theories, 
the space of massive $\qt$-BPS states $\widehat{\HHH}^{\rm BPS}$ 
is larger at every level than the 
space $\HHH^{\rm BPS}$ of generic massive $\qt$-BPS states, 
as the conformal field theoretic elliptic genus  counts BPS states with signs. 
Ultimately, we wish to know to what extent
one can identify the very states in the class of $\Zt$-orbifold 
CFTs, whose contributions cancel in the net count 
of $\EEE$ under selected deformations. The better we understand  them, in particular the type of 
group action they may enjoy, the more we can hope to uncover the 
VOA structure(s) on the generic space of states that 
\emph{does} contribute to the net count.
Our guiding principle in this quest is symmetry surfing \cite{taormina2013overarching,taormina2015symmetry}, a 
programme we have developed over a period of years and that has passed a number 
of non-trivial tests, either through explicit calculations within $\Zt$-orbifold 
CFTs \cite{taormina2015twist} or through a process of deformations away from the $\Zt$-orbifold 
point in two very interesting papers \cite{gaberdiel2017mathieu,keller2019lifting}.

In \cite{taormina2015twist}, symmetry surfing identifies a special 
one-dimensional subspace of the 16-dimensional space of twisted ground 
states in $\Zt$-orbifold conformal field theories on K3. Indeed  in such theories, corresponding to the $16$ fixed 
points of the standard $\Z_2$ action on $T_\Lambda$,  there are $16$ pairwise orthogonal  twisted
ground states $|\alpha_\beta\rangle$, labelled by $\beta\in\F_2^4$. By construction, the ``diagonal'' state
\be
|\alpha_{\rm diag}\rangle := \sum_{\beta\in \F_2^4}\,|\alpha_\beta\rangle
\ee
is invariant under all symmetries induced by geometric symmetries of the torus
$T_\Lambda$, including shifts by elements of $\hf\Lambda$. It is thus invariant under 
the full overarching affine group
$$
G:={\rm Aff}(\F_2^4)=\Zt^4 \rtimes GL(\F_2^4)
\stackrel{\mbox{\scriptsize\cite{jordan1870traite}}}{\cong} \Zt^4 \rtimes A_8
$$
which contains all the  groups $G_i$ of  finite symplectic automorphisms on Kummer surfaces. 

Conveniently, in \cite{taormina2015twist} we found that at massive level one, the orthogonal complement
of the Fock space built on $|\alpha_{\rm diag}\rangle$
echoes the construction of a 45-dimensional representation of the group $M_{24}$ 
by Margolin \cite{margolin1993geometry}. Inspired by his construction, we 
 thus define $\HHH^\perp\subset\wh\HHH^\bps$ as the 
Fock space built on the 15-dimensional  orthogonal 
complement of $|\alpha_{\rm diag}\rangle$ in the space of twisted ground states. 
This prompts the following ansatz\footnote{All direct sums are understood
as orthogonal direct sums.}, introduced similarly in \cite{gaberdiel2017mathieu},
\be\label{ansatz}
\wh\HHH^{\bps}=\HHH^\perp \oplus \HHH^{{\rm rest}}\oplus \HHH^+
\ee
\mbox{with }
$\HHH^\perp=\bigoplus\limits_{n=1}^{\infty}\,(H_n^\perp \otimes {\cal H}_n^{N=4})$,
$\HHH^{{\rm rest}}=\bigoplus\limits_{n=1}^{\infty}\,(H_n^{{\rm rest}} \otimes {\cal H}_n^{N=4})$,
$\HHH^+ =\bigoplus\limits_{n=1}^{\infty}\,(H_n^+ \otimes {\cal H}_n^{N=4})$, where as representations
of $\qu J_0$ and the octad group $G$,
\be\label{multspaces}
H_n\cong H_n^\perp \oplus H_n^{{\rm rest}}\qquad \mbox{ for all }
n\in \N, n>0,
\ee
and where $H_n$ was defined in \eqref{HBPS}.
Ultimately, \eqref{multspaces} should extend to an isomorphism of representations 
of $\qu J_0$ and $M_{24}$. As was pointed out in \cite{gaberdiel2017mathieu},
the ansatz \eqref{ansatz} can be interpreted  as identifying the massive contributions
to the generic space of
states as the subspace $\HHH^\perp \oplus \HHH^{{\rm rest}}$
of $\wh\HHH^\bps$ which remains at the BPS bound under
deformations of our CFT by the exactly marginal deformation
$T_{\rm diag}$ built on $|\alpha_{\rm diag}\rangle$.

Table \ref{table:table1} summarises data for the first four conformal 
weights above threshold in terms of the Fourier coefficients of the generating functions 
$ f(\tau)$, $g^{\rm inv}(\tau)$ and $ g^{\rm tw}(\tau)$ 
for massive $\qt$-BPS ground states contributing to the partition functions 
$U_{\ell=\hf}, U_{\ell=0}$ and $T_{\ell=0}$ respectively.
\begin{table}
\small{
\begin{tabular}{|c|cccc|cl|}
\hline
&&&&&&\\[-10pt]
\scriptsize{level $n$}&\scriptsize{1}&\scriptsize{2}&\scriptsize{3}&\scriptsize{4}&&\\[2pt]
\hline
&&&&&&\\[-5pt]
\scriptsize{$A_n$}&\scriptsize{90}&\scriptsize{462}&\scriptsize{1540}&\scriptsize{4554}&&\scriptsize{net number of states in $\EEE^{\NS}$}\\[5pt]
\scriptsize{$f_n$}&\scriptsize{3}&\scriptsize{1}&\scriptsize{18}&\scriptsize{15}&&\scriptsize{untwisted sector ($U_{\ell=\hf}$)}\\
\scriptsize{$ g^{\rm inv}_n$}&\scriptsize{0}&\scriptsize{16}&\scriptsize{8}&\scriptsize{72}&&\scriptsize{untwisted sector ($U_{\ell=0}$)}\\[5pt]
\scriptsize{$g^{\rm tw}_n$}&\scriptsize{6}&\scriptsize{28}&\scriptsize{98}&\scriptsize{282}&&\scriptsize{one twisted sector ($T_{\ell=0}$)}\\\
&&&&&&\\
\scriptsize{$A_n=16 g^{\rm tw}_n+g^{\rm inv}_n-2 f_n$}
&\scriptsize{96-6}&\scriptsize{448+16-2}&\scriptsize{1568+8-36}&\scriptsize{4512+72-30}&&\scriptsize{dim ${H}_n$ ($\HHH^{\bps}$)}\\[5pt]
\scriptsize{$g^{\rm tw}_n+ g^{\rm inv}_n-2f_n$}
&\scriptsize{0}&\scriptsize{42}&\scriptsize{70}&\scriptsize{324}&&\scriptsize{dim $H_n^{{\rm rest}}$ ($\HHH^{{\rm rest}}$)}\\[5pt]
\scriptsize{$15g^{\rm tw}_n$}&\scriptsize{90}&\scriptsize{420}&\scriptsize{1470}&\scriptsize{4230}&&\scriptsize{dim $H_n^\perp$ ($\HHH^\perp$)}\\[7pt]
\scriptsize{$\widehat{A}_n=16g^{\rm tw}_n+g^{\rm inv}_n+2 f_n$}
&\scriptsize{96+6}&\scriptsize{448+16+2}&\scriptsize{1568+8+36}&\scriptsize{4512+72+30}&&\scriptsize{dim $\widehat{H}_n$ ($\widehat{\HHH}^{\bps}$)}\\[3pt]
\scriptsize{$g^{\rm tw}_n+g^{\rm inv}_n+2 f_n$}&\scriptsize{12}&\scriptsize{46}&\scriptsize{142}&\scriptsize{384}&&\scriptsize{dim $\widehat{H}_n^{{\rm rest}}$}\\[3pt]
\hline
\end{tabular}}
\caption{\small{Data on the number of $\qt$-BPS states emerging from different sectors of $\Zt$-orbifolds CFTs
on K3.}} \label{table:table1}
\end{table}
By construction, we have $\dim H_n=A_n$ and
$\dim H_n^\perp=15g_n^{\rm tw}$ and hence, by \eqref{Adeco},
$\dim H_n^{{\rm rest}}= g^{\rm tw}_n+g^{\rm inv}_n-2 f_n$  
for all $n\in\N$, $n>0$. In other words, the excess states in $H_n^+$, which are lifted from
the BPS bound under a deformation by $T_{\rm diag}$, 
belong to
the twisted sector generated by
$|\alpha_{\rm diag}\rangle$ and the untwisted sector. While the results
of \cite{song2017chiral,wendland2019hodge} imply that the holomorphic untwisted states accounted for by
$2f_n$ all belong to $H_n^+$, it is not possible at this stage to identify 
whether the  remaining states in $H_n^+$
come from the diagonal twisted sector,
the untwisted sector, or both. We will return to this point in Section \ref{section:3}. 
\subsection{$\qt$-BPS states at level one and two}
To investigate the elusive properties of $H_n^+$ in general, we begin by studying the spaces
$\widehat{H}_n = H_n^\perp 
\oplus H_n^{\rm rest}\oplus H_n^+$ at levels $n=1$ and $n=2$ more closely. 
We also introduce a consistent action of $\geo\times\lgeo$ on $H_n^{\rm rest}\oplus H_n^+$
at these levels.
The rationale behind our construction will be explained in Section \ref{section:3} -- based
on the data collected at levels $n=1$ and $n=2$.
Here and in the 
following, we denote the modes of the four free fermions and of their superpartners as 
$(\chi_\pm^k)_\ell$ and $(a_\pm^k)_m$ with $\ell$ and $m$ either integers or half-integers in 
accordance with the boundary conditions imposed by the $\Zt$-orbifold construction. Following
a wide-spread tradition, $\ell$ and $m$ account for the \emph{negative} contributions to the energy. \\
\newpage
\underline{\textbf{Level 1}}\\
At conformal weight $n=1$,  since $A_1=90=15g_1^{\rm tw}$ and
$g_1^{\rm inv}=0$,  our ansatz is compatible with
the claim that the contributions to $\EEE^\NS$ from the six-dimensional  ($2f_1=6$) space of untwisted
massive $\qt$-BPS states cancels those from the six-dimensional
($g_1^{\rm tw}=6$) space of massive $\qt$-BPS states in the diagonal twisted sector. This 
was already discussed in \cite{taormina2015twist}, where the geometric action 
of the group $\geo$  was  mentioned. We reproduce our results here in the 
Neveu-Schwarz sector, not the least because some interesting lessons can be drawn from this case.

To work in the Neveu-Schwarz sector of the $\Zt$-orbifold CFTs, we spectral flow 
from the  Ramond sector and choose  
chiral-chiral  ground states, i.e. states in the kernel of $(2L_0-J_0)$ and $(2\qu L_0-\qu J_0)$.

In the subspace of the untwisted sector  accounted for by the partition 
function $U_{\ell=\hf}(z)$  (see \eqref{Uhf}), this amounts to building states from 
the bosonic highest weight states
$\Omega$ with conformal weights $(h, \qu{h})=(0,0)$ and 
$\widecheck\Omega = (\qu{\chi}_+^1)_{-\hf} (\qu{\chi}_+^2)_{-\hf}\Omega$ 
with  $(h, \qu{h})=(0,1)$ (recall  that $J_0=2J^3_{0}$  in \eqref{N4}).
The six-dimensional space of
states contributing to $2f_1=6$ is generated by the following 
$N=4$ primaries:
\be\label{leveloneuntw}
\begin{array}{l}\displaystyle
\kop_{-\hf}\ktm_{-\hf}\Omega,
\quad
\{\kop_{-\hf}\kom_{-\hf}-\ktp_{-\hf}\ktm_{-\hf}\,\}\Omega,
\quad
\ktp_{-\hf}\kom_{-\hf}\Omega,\\[5pt]
\ds
\kop_{-\hf}\ktm_{-\hf} \widecheck\Omega ,\quad
\{\kop_{-\hf}\kom_{-\hf}-\ktp_{-\hf}\ktm_{-\hf}\} \widecheck\Omega ,\quad 
\ktp_{-\hf}\kom_{-\hf}\widecheck\Omega.
\end{array}
\ee
The vacuum $\Omega$ is by definition invariant under $\geo \times \overline{SU(2)}_{{\rm geom}}$ 
and so is $\widecheck\Omega$, since $\qu{\chi_+^a}$ with $a \in \{1,2\}$ is a doublet under  
$\overline{SU(2)}_{{\rm geom}}$ and a singlet under $\geo$. Hence each row in \eqref{leveloneuntw}
generates  a  $(3,1)$ representation of the group 
$\geo \times \overline{SU(2)}_{{\rm geom}}$.
In the diagonal twisted sector, the six dimensional space of states accounted for by 
$g_1^{\rm tw}=6$  is built on the 
$\geo\times \overline{SU(2)}_{{\rm geom}}$ invariant twisted ground state $|\alpha_{{\rm diag}}\rangle$ and also 
transforms as the sum of two $(3,1)$ representations of the $\geo\times \overline{SU(2)}_{{\rm geom}}$ 
action,
generated by
\be
\begin{array}{l}\displaystyle
\aop_{-\hf}\ktm_0|\alpha_{{\rm diag}}\rangle,\quad
\{\aop_{-\hf}\kom_{0}-\atp_{-\hf}\ktm_{0}\}|\alpha_{{\rm diag}}\rangle,\quad
\atp_{-\hf}\kom_0|\alpha_{{\rm diag}}\rangle, \\[5pt]
\aom_{-\hf}\kom_0|\alpha_{{\rm diag}}\rangle,\quad
 \{\aom_{-\hf}\ktm_{0}+\atm_{-\hf}\kom_{0}\}|\alpha_{{\rm diag}}\rangle,\quad
  \atm_{-\hf}\ktm_0|\alpha_{{\rm diag}}\rangle.
\end{array}
\ee
So at level 1, the ``excess'' $\qt$-BPS states  in $H^+$
belong to a pair of isomorphic representations of $\geo\times \overline{SU(2)}_{{\rm geom}}$ of opposite fermion
number. A detailed analysis of the fate of this 12-dimensional space of states 
when the K3 theories are deformed away from the $\Zt$-orbifolds
has recently been carried out in \cite{keller2019lifting}. 
The conclusion is that under the diagonal deformation $T^{\rm diag}$, these states combine into  non-BPS 
representations and thus cease to be accounted for by 
$\EEE^{\NS}$; in other words,  deformations of a non-generic theory `lift' a 
number of massive $\qt$-BPS states, including all those
contributing to $U_{\ell=\hf}(z)$. That this must happen 
under deformations to generic theories follows already
from the analysis of \cite{wendland2019hodge}. \\

\underline{\textbf{Level 2}}\\
At conformal weight $n=2$,  in $H_2^+$
there is only a two-dimensional space of untwisted  
holomorphic excess states 
counted with one sign ($2f_2=2$). A priori we must   
find a two-dimensional subspace of 
the space $H_2^{\rm rest}\oplus H_2^+$ of dimension
$g^{\rm inv}_2+g^{\rm tw}_2=44$, contributing to $\EEE^\NS$ 
with the opposite sign, in order to identify the subspace of $H_2^+$  
matching  the untwisted holomorphic excess
states. The latter are singlets of $\geo\times\lgeo$ given by
\be \label{singletsuntwistedB2}
\begin{array}{rcl}\ds
|s_{(2)}\rangle&:=&\ds
\Big\{\sum_{k=1}^2(a_+^k)_{-1}(a_-^k)_{-1}
-\sum_{k=1}^2\{(\chi_+^k)_{-3/2}(\chi_-^k)_{-{1\over2}}-(\chi_+^k)_{-{1\over2}}(\chi_-^k)_{-3/2}\}\\[5pt]
&&\ds\phantom{spaceseekerspaceseeker}+2(\chi_+^1)_{-{1\over2}}(\chi_-^1)_{-{1\over2}}(\chi_+^2)_{-{1\over2}}(\chi_-^2)_{-{1\over2}}\Big\}
\ds \Omega ,\\[5pt]
|\widecheck s_{(2)} \rangle
&:=&\ds\Big\{\sum_{k=1}^2(a_+^k)_{-1}(a_-^k)_{-1}
-\sum_{k=1}^2\{(\chi_+^k)_{-3/2}(\chi_-^k)_{-{1\over2}}-(\chi_+^k)_{-{1\over2}}(\chi_-^k)_{-3/2}\}\\[5pt]
&&\ds\phantom{spaceseekerspaceseeker}+2(\chi_+^1)_{-{1\over2}}(\chi_-^1)_{-{1\over2}}(\chi_+^2)_{-{1\over2}}(\chi_-^2)_{-{1\over2}}
\Big\}\widecheck\Omega.
\end{array}
\ee
Remarkably, there are exactly two massive $\qt$-BPS singlets  with respect to the 
$\geo\times \overline{SU(2)}_{{\rm geom}}$ action
in the diagonal
twisted sector to match $|s_{(2)}\rangle$ and $|\widecheck s_{(2)}\rangle$. 
These are also the only singlets under the action of the diagonal $SU(2)$ in $\geo\times\lgeo$.
There are none in the untwisted sector accounted for by
$U_{\ell=0}$, a fact which serves as a 
justification 
in Section \ref{section:3} when we generalise our arguments. 
With the notation
\be \label{Anotation}
A_{\epsilon \delta \rho}^{k\ell mn}:=(a^k_{\epsilon})_{-\frac{1}{2}}\,(a^\ell_{\delta})_{-\frac{1}{2}}\,(a^m_{\rho})_{-\frac{1}{2}}\,(\chi^n_-)_0,\;\; \epsilon,\delta, \rho \in \{+,-\},\quad k,\ell,m,n \in \{1,2\},\;\;
\widecheck k:=3-k,
\ee
the two $\geo\times \overline{SU(2)}_{{\rm geom}}$ singlets in the diagonal twisted sector are given by
\be \label{singletstwistedD2}
\begin{array}{rcl}
\ds
|\tilde{s}_{(2)}\rangle
&:=&\ds
\sum_{k=1}^2\left\{\,(a_+^k)_{-\frac{3}{2}}(\chi_-^k)_0-3(a_+^k)_{-\hf}(\chi_-^k)_{-1}-3(-1)^k(a_+^k)_{-\hf}(\chi_+^{\widecheck{k}})_{-1}(\chi_-^1)_0(\chi_-^2)_0\right.\\[5pt]
&&\ds
\left.\phantom{spacechaserchaser}+2(A_{++-}^{kkkk}+A_{++-}^{k\widecheck{k}\widecheck{k}k})\right\}\,|\alpha_{{\rm diag}}\rangle,\\[5pt]
\ds
|\widecheck{\tilde{s}}_{(2)}\rangle
&:=&
\ds\sum_{k=1}^2(-1)^k\,\left\{\,(a_-^k)_{-\frac{3}{2}}(\chi_-^{\widecheck{k}})_0-3(a_-^k)_{-\hf}(\chi_-^{\widecheck{k}})_{-1}+3(-1)^k(a_-^k)_{-\hf}(\chi_+^k)_{-1}(\chi_-^1)_0(\chi_-^2)_0\right.\\[5pt]
&&\ds\left.\phantom{spacechaserchaser}+2(A_{+--}^{kkk\widecheck{k}}
-A_{+--}^{k\widecheck{k}kk})\right\}\,|\alpha_{{\rm diag}}\rangle.\\[5pt]
\end{array}
\ee
The remaining space of massive level two $\qt$-BPS states
 in the diagonal twisted sector is $26$-dimensional
($g^{\rm tw}_2=28$) and is  presented in Appendix 
\ref{subappendix:twistedstates}, while the 16-dimensional space of massive $\qt$-BPS states 
 in the untwisted sector accounted for by
$U_{\ell=0}(z)$ is presented in  Appendix \ref{subappendix:untwistedstates}. This detailed analysis 
supports the conjecture that the matching of excess $\qt$-BPS states respects the $\geo$ 
and the $ \overline{SU(2)}_{{\rm geom}}$ actions on 
these states, both in the untwisted and twisted sectors. 
We thus expect that the four states $|s_{(2)}\rangle$, $|\widecheck s_{(2)}\rangle$, $|\tilde s_{(2)}\rangle$, 
$|\widecheck{\tilde{s}}_{(2)}\rangle$ are lifted from the BPS bound under a deformation by $T_{\rm diag}$. It would
be interesting to confirm this prediction by 
conformal perturbation methods along the lines of \cite{keller2019lifting}.
In the next section we present general results corroborating our  conjecture at low levels.

\section{Geometric $SU(2)$ as a guiding principle }\label{section:3}
Let us now give a general definition of the $\geo\times\lgeo$ action on the 
space $H^{\rm rest}\oplus H^+$, along the lines indicated
 in Section \ref{subsection:freefermions}. We have defined this space as a 
 space of massive ground states
common to all $\Zt$-orbifold conformal field theories on K3, containing both twisted and
untwisted contributions, mindful however that its  decomposition into 
$H^{\rm rest}\oplus H^+$ has not been carried out so far.
All states in $H^{\rm rest}\oplus H^+$ are elements of 
the Fock space representations obtained from the vacuum $\Omega$ and the diagonal
twisted ground state $|\alpha_{\rm diag}\rangle$ by the action of the modes of 
the free fermionic fields $\chi_\pm^a, a\in \{1,2\}$, and their superpartners 
$j_\pm^a$, along with their antiholomorphic analogues. It thus suffices to state the action of 
$\geo\times\lgeo$ on these fields, on the vacuum $\Omega$ and on $|\alpha_{\rm diag}\rangle$,
yielding an action on the entire Fock space built on these states.
For each $\Zt$-orbifold conformal field theory on K3 we do this by using the left- and
the right-moving action of the group $SU(2)$ which contains the 
linear part of the geometric symmetry
group of our theory. In other words, we use the standard action of the group $SU(2)$ on 
$\mathbb{C}^2$, which is a subgroup of a global $SO(4)$ symmetry group of our 
$N=4$ superconformal algebra, see for instance \cite[\S5.3]{david2002microscopic}.
Both $\Omega$ and $|\alpha_{\rm diag}\rangle$
are invariant under every geometric symmetry group for any $\Z_2$-orbifold conformal
field theory on K3. We therefore choose both these states
to be invariant under $\geo\times\lgeo$, while
the fields $\chi_\pm^a$ and  $j_\pm^a$,
$a\in \{1,2\}$,
 transform as doublets under the action of $\geo$ and trivially under $\lgeo$.
Note that these fields also carry  a  $U(1)$ charge 
 associated with the  affine subalgebra $\frak{su}(2)$ of the $N=4$ superconformal 
 algebra. Our conventions for the two sets of charges, which are summarised 
in Table \ref{tab:charges},
 inform on how  to  refine the holomorphic partition functions for 
 the untwisted and twisted sectors so they
encode the action of the group $\geo$ on $\qt$-BPS states. 
The charges of $\overline{\chi^a_\pm}$ and $\overline{j^a_\pm}$ under 
$\overline{SU(2)}_{{\rm geom}}$ and $\overline{\frak{su}(2)}$ are  analogous.
From our derivation of
the $\geo\times\lgeo$ action on $H^{\rm rest}\oplus H^+$, we do not expect any meaningful extension to $H^\perp$. 
Indeed, 
although a  well-defined action of $SU(2)$ exists on $H^\perp$, which is trivial 
on every twisted ground state, this action does not extend the action of the linear
parts of our geometric symmetry groups to $SU(2)$, since these groups
act non-trivially on $H^\perp$.

\begin{table}[ht]
\begin{center}
\begin{tabular}{|c|c|c|}\hline
\diagbox{$\frak{su}(2)$}{$\geo$}&$+1$&$-1$\\
\hline
&&\\[-10pt]
$+1$&$\chi_+^1$& $\chi_+^2$\\[5pt]
$-1$&$\chi_-^2$&$\chi_-^1$\\
\hline
\end{tabular}
\qquad
\begin{tabular}{|c|c|c|}\hline
\diagbox{$\frak{su}(2)$}{$\geo$}&$+1$&$-1$\\
\hline
&&\\[-10pt]
$0$&$j_+^1$& $j_+^2$\\[5pt]
$0$&$j_-^2$&$j_-^1$\\
\hline
\end{tabular}
\caption{Charges of free bosons and fermions under $SU(2)_{\rm geom}$ and under the affine $\frak{su}(2)$ subalgebra of the $N=4$ superconformal algebra.}\label{tab:charges}
\end{center}
\end{table}
We introduce the complex variable
$w:=e^{2\pi i \nu}$, $\nu \in \C$, to track the $\geo$ charges of $\qt$-BPS states 
by refining the partition functions $U_{\ell={\hf}}$, $U_{\ell=0}$, 
$T_{\ell=0}$ introduced in  \eqref{Uhf}, \eqref{U0} and \eqref{T0} to
\begin{subequations}
\ba
U_{\ell=\hf}(z,\nu)&:=& -\frac{1}{2}(w^{-1}-2+w)\,\displaystyle{\frac{\vartheta_3(z+\nu)\,\vartheta_3(z-\nu)}{\vartheta_1(\nu)^2}}+\frac{1}{2}(w^{-1}+2+w)\,\displaystyle{\frac{\vartheta_4(z+\nu)\,\vartheta_4(z-\nu)}{\vartheta_2(\nu)^2}},\label{Uhalftheta}\nonumber\\
&&\\
U_{\ell=0}(z,\nu)&:=&-(w^{-1}-2+w)\,\frac{\vartheta_3(z+\nu)\,\vartheta_3(z-\nu)}{\vartheta_1(\nu)^2}-(w^{-1}+2+w)\,\frac{\vartheta_4(z+\nu)\,\vartheta_4(z-\nu)}{\vartheta_2(\nu)^2},\nonumber\\
&&\label{Uzerotheta}\\
T_{\ell=0}(z, \nu)&:=&\hf \frac{\vartheta_2(z+\nu)\,\vartheta_2(z-\nu)}{\vartheta_4(\nu)^2}-\hf \frac{\vartheta_1(z+\nu)\,\vartheta_1(z-\nu)}{\vartheta_3(\nu)^2}\label{Tzerotheta}.
\ea
\end{subequations}
Given the  action of $\overline{SU(2)}_{{\rm geom}}$ introduced above,  
the only states accounted for by the above partition functions which carry a non-trivial
action of this group are those built on the $\Zt$-orbifold odd ground states  
$\widetilde{\Omega}^1:=(\qu{\chi_+^1})_{-\hf}\,\Omega$ 
and $\widetilde{\Omega}^2:=(\qu{\chi_+^2})_{-\hf}\,\Omega$.
$U_{\ell=0}$ is the graded character for the space of
$\Zt$-orbifold invariant states in the  Fock space built on these two states, which
 transform
as a doublet under $\lgeo$. 
To encode this action as well, we therefore 
multiply $U_{\ell=0}(z,\nu)$ by $\hf\left(\qu w+\qu w^{-1}\right)$. 
We treat $\qu w$ as a formal variable separately from $w$, to keep the actions
of $\geo$ and $\lgeo$ apart, in the spirit of separating the action of the left- and the 
right-moving $N=4$ superconformal algebras. 

The action of the diagonal $SU(2)$ in
$\geo\times\lgeo$ is then captured by identifying $\qu w$ as the complex conjugate
of $w$. Indeed, by what was said above, for any given K3 theory the
action of $\geo$ and $\lgeo$ induces the action of the  linear part of the
geometric symmetry group $G_i$ mentioned in Section \ref{subsection:decomposition}
on holomorphic and antiholomorphic fields, respectively.
If $G_i$ acts on the holomorphic fields by the representation 
$\varrho$, then it acts by the complex conjugate representation $\qu\varrho$ 
on the antiholomorphic partner fields. This is used, for example, in the construction 
of the corresponding partition functions, where the contributions from
the antiholomorphic fields are simply obtained as the complex conjugates
of the contributions coming from their holomorphic partners (see, for
example, \cite[(5.2)]{eguchi1989superconformal}).
We remark that  the partition functions  $U_{\ell={\hf}}(z,\nu)$, 
$\hf\left(\qu w+\qu w^{-1}\right)U_{\ell=0}(z,\nu)$, $T_{\ell=0}(z,\nu)$
were used in \cite[(C.5), (C.8), (C.10)]{gaberdiel2017mathieu} at 
$z=0$ and at three specific values of $w$, 
with complex conjugates $\qu w$, which were interpreted as eigenvalues of the 
elements $g$ of the linear parts of the
geometric symmetry groups $G_i$. 
In that situation, $\qu w=w^{-1}$, and the expressions in 
\cite[(C.5), (C.8), (C.10)]{gaberdiel2017mathieu}
are invariant under $\qu w\leftrightarrow w$.
Indeed, by construction, at these special values our partition 
functions are the
characters of $g$. In \cite{gaberdiel2017mathieu},  the latter play a crucial 
role in providing evidence for the symmetry surfing programme.  

Let us first restrict our attention to the action of $\geo$.
Since the $\geo$ action commutes with that of $N=4$
and using \eqref{Uhf}, \eqref{U0}, \eqref{T0}, the refined partition functions  
enjoy a decomposition in $N=4$ superconformal characters of the form 
\begin{subequations}
\ba
U_{\ell={1\over2}}(z,\nu)
&=&
\chi_0^{\NS}(z)+f(\nu)\widetilde{\chi}^{\NS}(z),\label{Uhalf}\\
U_{\ell=0}(z,\nu)&=&
2(w+w^{-1})\,\chi_\hf^{\NS}(z)+g^{{\rm inv}}(\nu)\widetilde{\chi}^{\NS}(z),\label{Uzero}\\
T_{\ell=0}(z,\nu)&=&\chi_\hf^{\NS}(z)+g^{{\rm tw}}(\nu)\,\widetilde{\chi}^{\NS}(z).\label{Tzero}
\ea
\end{subequations}
As before, and by abuse of notation, $f(\nu)$ is our shorthand notation for $f(\tau,\nu)$,
and $f(\tau)=f(\tau,\nu=0)$, etc.
By construction, $\geo$ maps the spaces of massive ground states that contribute to
$U_{\ell={1\over2}}$, $U_{\ell=0}$,  $T_{\ell=0}$, respectively, to themselves, such that 
these three spaces decompose into direct sums of irreducible representations of $\geo$. Therefore,
\begin{subequations}
\ba
f(\nu)
&=&
\sum_{n=1}^\infty q^n\,\left(\sum_p f_{n,p}\chi_p^{SU(2)}(\nu)\right),\label{Bnusu2}\\
g^{{\rm inv}}(\nu)
&=&
\sum_{n=1}^\infty q^n\,\left(\sum_p g^{{\rm inv}}_{n,p}\chi_p^{SU(2)}(\nu)\right),\label{Cnusu2}\\
g^{{\rm tw}}(\nu)
&=&
\sum_{n=1}^\infty q^n\, \left( \sum_p g^{{\rm tw}}_{n,p},\chi_p^{SU(2)}(\nu)\right),\label{Dnusu2}
\ea
\end{subequations}
with the $SU(2)$ character of the representation with isospin $p\in{1\over2}\N$ given by
\be\label{su2irreps}
\chi _p^{SU(2)}(\nu):=\sum_{r=0}^{2p}e^{2\pi i(2p-2r)\nu}=\frac{\sin 2\pi(2p+1)\nu}{\sin 2\pi \nu}
\ee
and $f_{n,p}$,  $g^{{\rm inv}}_{n,p}$ and $g^{{\rm tw}}_{n,p}\in \mathbb{N}$ 
the multiplicity of the  $(2p+1)$-dimensional representation at level $n$ in the two untwisted 
and the twisted diagonal sectors, respectively.

In fact, as explained in Appendix \ref{subappendix:BCDhnu}, 
using identities amongst Jacobi theta functions and Appell functions, one may rewrite  
\eqref{Bnusu2}--\eqref{Dnusu2}  as
\begin{subequations}
\ba
f(\tau, \nu)
&=&-1-\big\{ \frac{1}{2}(w-2+w^{-1})\,h_1(\tau, \nu)
- \frac{1}{2}(w+2+w^{-1})\,h_2(\tau, \nu)\big\}\,\eta(\tau)\,q^{1\over8},
\qquad\label{Uhffinal}\\
g^{{\rm inv}}(\tau, \nu)
&=&-\big\{ (w-2+w^{-1})\,h_1(\tau, \nu)+(w+2+w^{-1})\,h_2(\tau, \nu)\big\}\,
\eta(\tau)\,q^{1\over8},\label{U0final}\\
g^{{\rm tw}}(\tau, \nu)
&=&-\hf\,\{h_3(\tau, \nu)+h_4(\tau, \nu)\,\}\,\eta(\tau)\,q^{1\over8}\label{T0final}.
\ea
\end{subequations}

The three functions $f(\tau,\, \nu)$, $g^{{\rm inv}}(\tau,\, \nu)$ 
and $g^{{\rm tw}}(\tau,\,\nu)$ are the graded characters of certain
subspaces of the space of massive $\qt$-BPS ground states, whose decomposition into 
$H^{\rm rest}\oplus H^+$ is at the heart of our
investigation. Recall that we have defined the space $H^+$ as to contain pairs of ground states at opposite fermion
numbers whose contributions to the  elliptic genus $\EEE^\NS$ cancel. Upon deformation 
of our K3 theory by $T^{\rm diag}$,
each such pair is lifted to a common long $N=4$ representation off the BPS bound. 
On the other hand,  for every 
$\Zt$-orbifold CFT on K3, $\geo\times\lgeo$ acts on $H^{\rm rest}\oplus H^+$
as to restrict to the action of
the linear part of the geometric symmetry group of the theory, 
which remains unbroken under deformations by $T^{\rm diag}$. This, together with the symmetry
surfing proposal, prompts us to 
postulate that the states in $H^+$ are paired up according to their 
transformation properties under $\geo\times\lgeo$. In other words, we postulate that $H^+$ decomposes
into pairs of isomorphic representations under $\geo$ and $\lgeo$ with opposite fermion numbers.

To determine which subspaces of the spaces accounted for by 
$f(\nu)$, $g^{\rm inv}(\nu)$ and $g^{\rm tw}(\nu)$ may contribute to $H^+$, we recall from the 
above that by the results of \cite{wendland2019hodge}, all untwisted holomorphic states accounted
for by $f(\nu)$ are non-generic and thus belong to $H^+$. All of them are bosonic, while 
$g^{\rm inv}(\nu)$ and $g^{\rm tw}(\nu)$ account for fermionic states only. In other words,
we must find $\geo\times\lgeo$ representations matching those accounted for by $f(\nu)$
within the spaces whose graded characters are $g^{\rm inv}(\nu)$ and $g^{\rm tw}(\nu)$.
Observe that  by \eqref{halfshiftinh}
and \eqref{Uhffinal}--\eqref{T0final}, $f(\nu)$ and
$g^{\rm tw}(\nu)$ are
invariant under a shift of the variable $\nu$ by $\hf$, while 
$g^{{\rm inv}}(\nu +\hf)=-g^{{\rm inv}}(\nu)$. 
This shows that $g^{{\rm inv}}(\nu)$ only  accounts for
representations of $\geo$ with 
half-integer spin, while $f(\nu)$ and $g^{{\rm tw}}(\nu)$ 
only account for representations with integer spin. 
Moreover, all states accounted for by $f(\nu)$ and $g^{\rm tw}(\nu)$ transform trivially
under $\lgeo$, in contrast to those accounted for by $g^{\rm inv}(\nu)$.
This, together with the evidence 
provided by the explicit calculations at levels $n=1, 2$,  prompts us to postulate that, at 
any level $n \in \N, n \ge 1$,  only $\qt$-BPS states belonging to the diagonal twisted 
sector  can  pair up with those occurring in the untwisted sector and accounted for by
$f(\nu)$.
In light of the decompositions \eqref{Bnusu2}--\eqref{Dnusu2}
into characters of $\geo$, this implies the claim that
\be\label{positivityclaim}
\forall p,\,n\in\N\colon\qquad
g^{{\rm tw}}_{n,p}-2f_{n,p}\ge 0.
\ee
We have expanded 
the  functions $f(\nu)$ and $g^{{\rm tw}}(\nu)$ as
$q$-power series   and verified \eqref{positivityclaim} up 
to $O(q^{101})$,
supporting our postulates. We present the data  up to $O(q^{16})$ in Table \ref{tab:datafouriercoef}
for reference.
We hope to provide an analytic proof of \eqref{positivityclaim} in the near future.

\section{Discussion}\label{discussion}

The VOA(s) underlying Mathieu Moonshine remain elusive to the extent that 
even a consensus on whether or not to expect a link between Mathieu Moonshine 
and  K3 theories has not been reached. The works of \cite{song2017chiral,wendland2019hodge} indicate
that such a link could involve a \emph{generic space of states} of K3 theories, in accordance with
the ideas behind our symmetry surfing programme \cite{taormina2010symmetries,taormina2013overarching,taormina2015twist,taormina2015symmetry}. The present
work is a contribution to the study of generic properties of K3 theories which we find interesting in
their own right. This can be viewed as a preparation for a new attempt at 
 the construction of a Mathieu Moonshine VOA on the generic space of states.

In \cite{wendland2019hodge}, crucially building on the results of \cite{song2017chiral}, it was shown that indeed there exists
a generic space of states $\H_0$ for all K3 theories, roughly defined through the property that it 
embeds into the space of BPS states of every K3 theory as a representation of the holomorphic $N=4$ superconformal
algebra, extended by the zero modes $\qu L_0,\, \qu J_0$ of the Virasoro field and
$\mathfrak{u}(1)$ current in the antiholomorphic $N=4$ superconformal algebra. Although $\H_0$
can be modelled by the chiral de Rham cohomology of a K3 surface \cite{song2017chiral,wendland2019hodge}, its detailed
structure has not been studied so far. Approaching the space from the perspective of \emph{non-generic}
yet accessible K3 theories, we have, in the present work, studied more closely the 
structure of the $\qt$-BPS 
states of  $\Zt$-orbifold CFTs of toroidal 
 theories. We propose a strategy to earmark the $\qt$-BPS states that
move off the BPS bound under the most symmetric deformation $T^{\rm diag}$ of $\Zt$-orbifold 
CFTs on K3, away from the $\Zt$-orbifold limit. Such states must come in pairs of opposite
fermion numbers, such that their contributions cancel each other in the 
conformal field theoretic elliptic genus of K3, in order to be part of the same 
(long) non-BPS representation after deformation. In \cite{taormina2015twist}, where the first concrete study of 
BPS states contributing to the count at level $n=1$ in the 
conformal field theoretic elliptic genus of K3
was undertaken, we 
identified a 15-dimensional space of twisted ground states 
carrying a Fock space representation $\HHH^\perp$ which turns
out to be generic in the above sense, that is, along deformations by $T^{\rm diag}$. In the twisted 
sector, $\HHH^\perp$ is the
orthogonal complement of a  `diagonal'  subspace 
from which $T^{\rm diag}$ arises. The findings were guided by the 
symmetry surfing programme \cite{taormina2013overarching,taormina2015symmetry} and were inspired by Margolin's construction of a 
$45$-dimensional representation of $M_{24}$ \cite{margolin1993geometry}. They highlighted the action of the octad 
group, a maximal subgroup of $M_{24}$, on the $45$-dimensional subspace of 
$\HHH^\perp$ at level 1. Evidence of the octad group action at all levels was provided in 
the work \cite{gaberdiel2017mathieu}, fueling the symmetry surfing programme whose aim is to exhibit an 
$M_{24}$ action on the generic space of states. More recently, Keller and 
Zadeh \cite{keller2019lifting} deformed the $\Zt$-orbifold  CFTs
away from the Kummer point by a marginal operator and showed that if the deformation is taken 
in the diagonal direction $T^{\rm diag}$, then indeed all the BPS states in the twisted sector
of the original non-generic K3  theory that are orthogonal to $\HHH^\perp$ move off
the BPS bound under the deformation.

It remains that beyond level $n=1$, we do not have total control on which BPS 
states move off the BPS bound
under a given deformation. By the results of \cite{wendland2019hodge}, we know that  under deformations
to generic K3 theories this happens for every untwisted massive state accounted for by the
partition function $U_{\ell=\hf}$. Since each state that moves off the BPS bound pairs up
with a state of opposite fermion number, to become part of the same long non-BPS representation, 
one needs to identify the correct partners  in either 
the untwisted subsector $U_{\ell=0}$ or the twisted  sector. 
Following the results of \cite{gaberdiel2017mathieu}, we know that under the deformation by
$T^{\rm diag}$, of the twisted states only those in the diagonal twisted sector can move
off the BPS bound. We postulate that none of the untwisted states accounted for in $U_{\ell=0}$
do.
At level $n=1$ this trivially holds as the  level one contribution to $U_{\ell=0}$
is zero. In fact, all states in the twisted diagonal sector  move off the BPS bound under deformation by $T^{\rm diag}$. 
This is a very special situation that does not persist at higher levels.

In order to identify potential states to pair up with  
the states accounted for in $U_{\ell=\hf}$ and
move off the BPS bound under deformation by $T^{\rm diag}$, we postulate compatibility with
a geometric action of the group $SU(2)$, denoted $\geo$ in this work. Indeed,
as already pointed out in \cite{taormina2015twist}, the holomorphic Dirac fermion fields 
$\chi _\pm^a, a \in \{1,2\}$, and their superpartners, which are the building blocks of  
$\Zt$-orbifold CFTs, transform as doublets under  $\geo$. 
The group acts
trivially on the vacuum and on the twisted diagonal ground state, and also on the antiholomorphic
partners of the $\chi _\pm^a, a \in \{1,2\}$, and their superpartners. We have introduced
refined partition functions that  keep track of that group action on $\qt$-BPS 
states  in the untwisted and twisted diagonal sectors, and we were able to 
show that at level $n\leq2$ only states stemming from the 
twisted diagonal sector carry representations of $\geo$ that match those in the untwisted subsector 
accounted for by $U_{\ell=\hf}$. The $\geo$ action on the $\qt$-BPS states thus helps 
to identify sets of states in the  
diagonal twisted sector that may lift off the BPS bound under the deformation by $T^{\rm diag}$. We note that a certain 
degree of indetermination remains, as the multiplicities of isospin $p$ representations at any fixed level 
$n$ in the diagonal twisted sector quickly exceed
by far twice the multiplicities of isomorphic representations at the same level in 
 the untwisted sector, as evidenced in Table \ref{tab:datafouriercoef}. 
Therefore, except for levels $n=1,2$, our postulate is not  powerful enough to pin down the 
exact states that are generic along the deformation by $T^{\rm diag}$.

As explained  above, $\geo$ in particular leaves the 
$\Zt$-orbifold odd ground states in the untwisted subsector accounted for by $U_{\ell=0}$ 
 invariant. From the perspective of symmetry surfing, it is perhaps more natural to consider the action 
of a diagonal $SU(2)$ in $\geo\times\lgeo$, where $\lgeo$ is the antiholomorphic analog of $\geo$. In
particular, under this diagonal action of $SU(2)$, the ground states in the untwisted sector accounted for
by $U_{\ell=0}$ transform non-trivially. We argue that representations from this
sector do not pair up with states accounted for by $U_{\ell=\hf}$
to form long representations off the BPS bound under any deformation. Indeed, if this were the case, then there should 
be a deformation in the underlying toroidal theory that would lift these states off the BPS bound. However, such states are generic to all 
toroidal theories and hence such deformations do not exist.

Altogether we  expect the role of $\geo$ 
to be helpful in understanding the generic space of states of K3 theories, independently of Mathieu 
Moonshine. An analysis of deformations beyond level $n=1$ along the lines of those followed
in \cite{keller2019lifting} would certainly shed more light on the relevance of $\geo$.
The analysis of \cite{keller2019lifting} already shows that any sufficiently small deformation away
from the $\Zt$-orbifold conformal field theories on K3 reduces the space of $\qt$-BPS states at massive
level one to a generic space. Given the results of \cite{wendland2019hodge}, the same must hold at
arbitrary level. It would be interesting to understand the structure of the underlying VOAs and
their dependence on the details of the deformation. Indeed, is it possible that the dependence
on the choice of deformation drops out entirely? Ultimately, this could answer some of the 
open questions of Mathieu Moonshine.
\appendix 
\section{Modular and mock modular input} \label{appendix:char}
\subsection{Jacobi Theta functions} \label{subappendix:jacobi}
Let  $q=e^{2\pi i \tau}, \tau \in \frak{H}$ and $y=e^{2\pi iz}, z \in \C$.
Our notations for the Jacobi theta functions are
\be
\begin{array}{rclcl}
\ds\vartheta_1(z)
&:=&\ds i\sum_{n=-\infty}^\infty\,(-1)^n\,q^{\hf(n-\hf)^2}\,y^{n-\hf}
&=&\ds iq^{1\over8}y^{-\hf}\prod_{n=1}^\infty\,(1-q^n)(1-q^{n-1}y)(1-q^ny^{-1}),\\
\ds\vartheta_2(z)
&:=&\ds\sum_{n=-\infty}^\infty\,q^{\hf(n-\hf)^2}\,y^{n-\hf}
&=&\ds q^{1\over8}y^{-\hf}\prod_{n=1}^\infty\,(1-q^n)(1+q^{n-1}y)(1+q^ny^{-1}),\\
\ds\vartheta_3(z)
&:=&\ds\sum_{n=-\infty}^\infty\,q^{\hf n^2}\,y^{n}
&=&\ds\prod_{n=1}^\infty\,(1-q^n)(1+q^{n-\hf}y)(1+q^{n-\hf}y^{-1}),\\
\ds\vartheta_4(z)
&:=&\ds\sum_{n=-\infty}^\infty\,(-1)^n\,q^{\hf n^2}\,y^{n}
&=&\ds\prod_{n=1}^\infty\,(1-q^n)(1-q^{n-\hf}y)(1-q^{n-\hf}y^{-1}),
\end{array}
\ee
with $\vartheta_i(0):=\vartheta_i, i\in \{2,3,4\}$ and $\vartheta_1(0)=0$.
All the theta function identities used in this paper can be found in \cite{whittaker1920course}. In particular, for $z_1,\,z_2\in \C$,
the following addition formulae are useful, all of which can be proved by residue analysis,
\be \label{jacobiidentities}
\begin{array}{rcccl}
\ds\vartheta_1(z_1+z_2)\vartheta_1(z_1-z_2)\,\vartheta_4^2
&=&\ds\vartheta_3(z_1)^2\vartheta_2(z_2)^2-\vartheta_2(z_1)^2\vartheta_3(z_2)^2
&=&\ds\vartheta_1(z_1)^2\vartheta_4(z_2)^2-\vartheta_4(z_1)^2\vartheta_1(z_2)^2,\\[4pt]
\ds\vartheta_4(z_1+z_2)\vartheta_4(z_1-z_2)\,\vartheta_2^2
&=&\ds\vartheta_4(z_1)^2\vartheta_2(z_2)^2+\vartheta_3(z_1)^2\vartheta_1(z_2)^2
&=&\ds\vartheta_2(z_1)^2\vartheta_4(z_2)^2+\vartheta_1(z_1)^2\vartheta_3(z_2)^2,\\[4pt]
\end{array}
\ee
and
\ba \label{jacobiidentities2}
\vartheta_2(z_1\pm z_2)\vartheta_3(z_1\mp z_2)\vartheta_2\vartheta_3
&=&\vartheta_2(z_1)\vartheta_3(z_1)\vartheta_2(z_2)\vartheta_3(z_2)
\mp\vartheta_1(z_1)\vartheta_4(z_1)\vartheta_1(z_2)\vartheta_4(z_2),\nonumber\\
\vartheta_2(z_1\pm z_2)\vartheta_4(z_1\mp z_2)\vartheta_2\vartheta_4
&=&\vartheta_2(z_1)\vartheta_4(z_1)\vartheta_2(z_2)\vartheta_4(z_2)
\mp\vartheta_1(z_1)\vartheta_3(z_1)\vartheta_1(z_2)\vartheta_3(z_2),\qquad\\
\vartheta_3(z_1\pm z_2)\vartheta_4(z_1\mp z_2)\vartheta_3\vartheta_4
&=&\vartheta_3(z_1)\vartheta_4(z_1)\vartheta_3(z_2)\vartheta_4(z_2)
\mp\vartheta_1(z_1)\vartheta_2(z_1)\vartheta_1(z_2)\vartheta_2(z_2).\nonumber
\ea
The Dedekind $\eta$ function is defined as
\be
\eta(\tau):=q^{1\over24}\,\prod_{n=1}^\infty (1-q^n),
\ee
and is related to the Jacobi theta functions by the identity
\be\label{strange}
2\eta^3=\vartheta_2\vartheta_3\vartheta_4.
\ee
For future reference, we note
\be
\vartheta_1^2(\frac{\tau+1}{2})=q^{-\qt}\,\vartheta_3^2(\tau), \quad
\vartheta_2^2(\frac{\tau+1}{2})=-q^{-\qt}\,\vartheta_4^2(\tau),\quad
\vartheta_3^2(\frac{\tau+1}{2})=0,\quad
\vartheta_4^2(\frac{\tau+1}{2})=q^{-\qt}\,\vartheta_2^2(\tau).
\ee
\subsection{Appell functions}\label{subappendix:appell}
Let 
\be
\KKK_\ell(\tau, \nu,\mu)
:=\sum_{m \in \Z}\frac{q^{\frac{\ell m^2}{2}}w^{\ell m}}{1-wxq^m},\quad w=e^{2\pi i \nu},\quad x=e^{2\pi i \mu}
\ee
be the Appell function at level $\ell \in \N$, $\nu, \mu \in \C, \nu+\mu \notin \Z\tau+\Z$ 
\cite{appell1884fonctions,semikhatov2005higher,zwegers2002mock}. Define
\be \label{h3}
h_3(\tau, \nu)
:=\frac{1}{\eta(\tau)}\frac{1}{\vartheta_3(\tau, \nu)}q^{-{1\over8}}\,
\KKK_1(\tau, \nu,-{\tau+1\over2})
=\frac{1}{\eta(\tau)}\frac{1}{\vartheta_3(\tau, \nu)}
\sum_{m \in \Z}\frac{q^{\frac{m^2}{2}-{1\over8}}w^{m}}{1+wq^{m-{1\over2}}}.
\ee
By evaluating \eqref{h3} at $\nu$ shifted by the three two-torsion points of 
an elliptic curve, one defines three new functions,
\be\label{h4h2h1}
\begin{array}{rclcl}
\ds h_4(\tau, \nu)
&:=&\ds h_3(\tau, \nu+\hf)
&=&\ds \frac{1}{\eta(\tau)}\frac{1}{\vartheta_4(\tau, \nu)}\sum_{m \in \Z}\frac{(-1)^m 
q^{\frac{m^2}{2}-{1\over8}}w^{m}}{1-wq^{m-{1\over2}}},\\[8pt]
\ds h_2(\tau, \nu)
&:=&\ds h_3(\tau, \nu+\frac{\tau}{2})
&=&\ds \frac{1}{\eta(\tau)}\frac{1}{\vartheta_2(\tau, \nu)}
\sum_{m \in \Z}\frac{q^{\frac{m^2}{2}+\frac{m}{2}}w^{m+{1\over2}}}{1+wq^{m}},\\[8pt]
\ds h_1(\tau, \nu)
&:=&\ds h_3(\tau, \nu+\frac{\tau +1}{2})
&=&\ds \frac{-i}{\eta(\tau)}\frac{1}{\vartheta_1(\tau, \nu)}
\sum_{m \in \Z}\frac{(-1)^mq^{\frac{m^2}{2}+\frac{m}{2}}w^{m+{1\over2}}}{1-wq^{m}}.
\end{array}
\ee
The following properties are immediate:
\be \label{propertieshfcts}
h_i(-\nu)=h_i(\nu),\qquad
h_i(\nu+\tau)=h_i(\nu)
\qquad\,\,\forall i \in\{1,2,3,4\}
\ee
and
\be\label{halfshiftinh}
h_1(\nu+\hf)=h_2(\nu),\quad
h_2(\nu+\hf)=h_1(\nu),\quad
h_3(\nu+\hf)=h_4(\nu),\quad
h_4(\nu+\hf)=h_3(\nu).
\ee

\underline{\textbf{Identities}}

Here and throughout the paper, $h_i:=h_i(\nu=0), i\in \{1,2,3,4\}$.
The following identities can be established by residue analysis,
\ba \label{identityresidue}
h_3(\nu)&=&h_3-\frac{\vartheta_1(\nu)^2}{\vartheta_3(\nu)^2}\,\frac{\eta^2}{\vartheta_3^2},\qquad h_4(\nu)=h_4+\frac{\vartheta_1(\nu)^2}{\vartheta_4(\nu)^2}\,\frac{\eta^2}{\vartheta_4^2},\qquad
h_2(\nu)=h_2+\frac{\vartheta_1(\nu)^2}{\vartheta_2(\nu)^2}\,\frac{\eta^2}{\vartheta_2^2}.\nonumber\\
\ea
New ones are obtained by shifting $\nu$ by the three two-torsion points in \eqref{identityresidue},
\ba \label{identityresidue2}
h_4(\nu)&=&h_3-\frac{\vartheta_2(\nu)^2}{\vartheta_4(\nu)^2}\,\frac{\eta^2}{\vartheta_3^2},\qquad h_3(\nu)=h_4+\frac{\vartheta_2(\nu)^2}{\vartheta_3(\nu)^2}\,\frac{\eta^2}{\vartheta_4^2},\qquad h_1(\nu)=h_2+\frac{\vartheta_2(\nu)^2}{\vartheta_1(\nu)^2}\,\frac{\eta^2}{\vartheta_2^2},\nonumber\\
h_2(\nu)&=&h_3+\frac{\vartheta_4(\nu)^2}{\vartheta_2(\nu)^2}\,\frac{\eta^2}{\vartheta_3^2},\qquad h_1(\nu)=h_4+\frac{\vartheta_4(\nu)^2}{\vartheta_1(\nu)^2}\,\frac{\eta^2}{\vartheta_4^2},\qquad  h_3(\nu)=h_2-\frac{\vartheta_4(\nu)^2}{\vartheta_3(\nu)^2}\,\frac{\eta^2}{\vartheta_2^2}, \nonumber\\
h_1(\nu)&=&h_3+\frac{\vartheta_3(\nu)^2}{\vartheta_1(\nu)^2}\,\frac{\eta^2}{\vartheta_3^2},\qquad h_2(\nu)=h_4+\frac{\vartheta_3(\nu)^2}{\vartheta_2(\nu)^2}\,\frac{\eta^2}{\vartheta_4^2},\qquad h_4(\nu)=h_2-\frac{\vartheta_3(\nu)^2}{\vartheta_4(\nu)^2}\,\frac{\eta^2}{\vartheta_2^2}.\nonumber\\
\ea
\subsection{$N=4$ characters at c=6}\label{subappendix:characters}
The  characters for unitary, irreducible representations of the $N=4$ 
superconformal algebra were first derived in \cite{eguchi1987unitary}. When the central charge is $c=6$, the 
representations fall into an infinite class of `long' or `massive'
representations with Neveu-Schwarz characters of the form
\be \label{massive}
q^h\,\widetilde{\chi}^{\NS}(\tau, z)=q^{h-{1\over8}}\,\frac{\vartheta_3(\tau,z)^2}{\eta(\tau)^3},
\ee
with $ h \in \R$, $h > 0$, the conformal weight of the highest weight state
alongside two `short' or `massless' representations labelled by the $\frak{su}(2)$ `spin' 
$\ell \in\{0, \hf \}$ and the conformal weight $h=\ell$ of their highest weight states. The corresponding  
Neveu-Schwarz characters are expressible in a variety of ways. Here, we 
use their expressions in terms of the Appell functions  $h_i(z)$ defined in \eqref{h3} and \eqref{h4h2h1}. 
We have
\be \label{charN4h3z}
\chi_{\hf}^\NS (\tau, z):=\chi_{h=\ell=\hf}^\NS (\tau, z)
=h_3(\tau,z)\,\frac{\vartheta_3(\tau,z)^2}{\eta(\tau)^2}
=h_3(\tau,z)\,\eta(\tau)\, q^{1\over8}\,\widetilde{\chi}^\NS(\tau, z).
\ee
This form of the characters was first presented in \cite{eguchi1988unitary} and 
expresses the branching of $N=4$ characters in an infinite sum of $N=2$ characters at central charge $c=6$.
Inserting \eqref{identityresidue} and \eqref{identityresidue2} 
into \eqref{charN4h3z}, one gets
\begin{subequations}
\ba
\chi_{\hf}^\NS (\tau, z)
&=&
-\frac{\vartheta_1(\tau,z)^2}{\vartheta_3(\tau)^2}
+h_3(\tau)\,\eta(\tau)\,q^{1\over8}\widetilde{\chi}^{\NS}(\tau, z) \label{charactersh13}\\
& =&\phantom{di}\frac{\vartheta_2(\tau,z)^2}{\vartheta_4(\tau)^2}
+h_4(\tau)\,\eta(\tau)\, q^{1\over8}\widetilde{\chi}^{\NS}(\tau, z)\label{charactersh24}\\
& =&-\frac{\vartheta_4(\tau,z)^2}{\vartheta_2(\tau)^2}
+h_2(\tau)\,\eta(\tau)\, q^{1\over8}\widetilde{\chi}^{\NS}(\tau, z).\label{charactersh42}
 \ea
 \end{subequations}

The second massless $N=4$ character in the  Neveu-Schwarz sector is given by
\be \label{magic}
\chi_0^\NS (\tau, z):=\chi_{h=\ell=0}^\NS (\tau, z)=\widetilde{\chi}^\NS(\tau, z)-2\chi_\hf^\NS (\tau, z).
\ee
Twisting by the fermion number operator one obtains
\be\label{flowtotwist}
\begin{array}{rcl}
\ds\widetilde{\chi}^{\widetilde\NS} (\tau, z) &=& \ds\widetilde{\chi}^\NS (\tau, z+\frac{1}{2}),\\[8pt]
\ds\chi^{\widetilde\NS}_\ell (\tau, z) &=& \ds\chi^\NS_\ell (\tau, z+\frac{1}{2}),\qquad
\ell\in\{0,\hf\}.
\end{array}
\ee
Under spectral flow, the Neveu-Schwarz and 
Ramond characters flow to each other as
\ba \label{sf}
\widetilde{\chi}^\NS (\tau, z+\frac{\tau}{2})
&=&\ds q^{-\qt}\,y^{-1}\,\widetilde{\chi}^{\,\Ra} (\tau, z),\quad
\widetilde{\chi}^\NS (\tau, z+\frac{\tau+1}{2})
=-q^{-\qt}\,y^{-1}\,\widetilde{\chi}^{\,\Rat} (\tau, z),
 \nonumber\\[5pt]
\chi_\ell^\NS (\tau, z+\frac{\tau}{2})
&=&\ds q^{-\qt}\,y^{-1}\,\chi_{\hf-\ell}^{\Ra} (\tau, z),\,
\chi_\ell^\NS (\tau, z+\frac{\tau+1}{2})
=-q^{-\qt}\,y^{-1}\,\chi_{\hf-\ell}^{\Rat} (\tau, z),\quad 
\ell \in\{ 0, \hf{}\}.\nonumber\\
\ea
Given \eqref{charN4h3z} and \eqref{magic}, all $N=4$ characters 
may be expressed in terms of Appell functions. In particular,
\be \label{charN4h2h4h1z}
\begin{array}{rclcl}
\ds\chi_{0}^{\Ra}(\tau, z)
&=&\ds h_2(\tau,z)\,\frac{\vartheta_2(\tau,z)^2}{\eta(\tau)^2}
&=&\ds h_2(\tau,z)\,\eta(\tau)\,q^{1\over8}\,\widetilde{\chi}^\Ra(\tau, z),\\[10pt]
 \ds\chi_{\hf}^{\widetilde{\NS}}(\tau, z)
 &=&\ds h_4(\tau,z)\,\frac{\vartheta_4(\tau,z)^2}{\eta(\tau)^2}
 &=&\ds h_4(\tau,z)\,\eta(\tau)\,q^{1\over8}\,\widetilde{\chi}^{\widetilde{\NS}}(\tau, z),\\[10pt]
 \ds\chi_{0}^{\Rat}(\tau, z)
 &=&\ds h_1(\tau,z)\,\frac{\vartheta_1(\tau,z)^2}{\eta(\tau)^2}
 &=&\ds h_1(\tau,z)\,\eta(\tau)\,q^{1\over8}\,\widetilde{\chi}^{\Rat}(\tau, z).
 \end{array}
\ee
The Witten index of the various representations is obtained by setting 
$z=0$ in the $\Rat$ characters. The massive representations all have Witten index zero while 
by \eqref{sf}, \eqref{magic} and \eqref{identityresidue},
\be
\begin{array}{rcccc}
\ds\chi_\hf^{\Rat}(\tau, 0)
&=&\ds -2 &=&\ds -q^\qt\,\chi_0^\NS (\tau, \frac{\tau+1}{2}),\\[5pt]
\ds\chi_0^{\Rat}(\tau, 0)
&=&\ds1&=&\ds -q^\qt\,\chi_\hf^\NS (\tau, \frac{\tau+1}{2}).
\end{array}
\ee

\subsection{The functions $f(\tau), g^{\rm inv}(\tau)$ and $g^{\rm tw}(\tau)$} \label{subappendix:BCDh}
The functions \eqref{BtauCtau} and \eqref{Dtau} can all be obtained by standard manipulations,
\ba
U_{\ell=\hf}(z)
&\stackrel{\eqref{Uhf}}{=}&
\hf\left\{\frac{\vartheta_3(z)^2}{\eta^6}+4\frac{\vartheta_4(z)^2}{\vartheta_2^2}\right\}\nonumber\\
&\stackrel{\eqref{massive},\eqref{charactersh42}}{=}&\frac{1}{2\eta^3}q^{1\over8}\widetilde{\chi}^{\NS}(z)+2\{ -\chi_\hf^{\NS}(z)+h_2\,\eta\, q^{1\over8}\widetilde{\chi}^{\NS}(z)\}\nonumber\\
&\stackrel{\eqref{magic}}{=}&
\chi_0^{\NS}(z)+\underbrace{\big\{2h_2\,\eta\,q^{1\over8}-1
+\frac{1}{2\eta^3}q^{1\over8}\big\}}_{=f(\tau)}\widetilde{\chi}^{\NS}(z),\ea
\ba
U_{\ell=0}(z)
&\stackrel{\eqref{U0}}{=}&
\frac{\vartheta_3(z)^2}{\eta^6}-4\frac{\vartheta_4(z)^2}{\vartheta_2^2}\nonumber\\
&\stackrel{\eqref{massive},\eqref{charactersh42}}{=}&
\frac{1}{\eta^3}q^{1\over8}\widetilde{\chi}^{\NS}(z)
-4\{ -\chi_\hf^{\NS}(z)+h_2\,\eta\, q^{1\over8}\widetilde{\chi}^{\NS}(z)\}\nonumber\\
&=&4\chi_\hf^{\NS}(z)+\underbrace{\{-4h_2\eta\,q^{1\over8}
+\frac{1}{\eta^3}q^{1\over8}\}}_{=g^{{\rm inv}}(\tau)}\widetilde{\chi}^{\NS}(z)
\ea
and
\ba\label{Tellzero}
T_{\ell=0}(z)
&\stackrel{\eqref{T0}}{=}&
\hf \left\{\frac{\vartheta_2(z)^2}{\vartheta_4^2}-\frac{\vartheta_1(z)^2}{\vartheta_3^2}\right\}\nonumber\\
&\stackrel{\eqref{charactersh13},\eqref{charactersh24}}{=}&
\hf \{\,\chi_\hf^\NS(z)-h_4\,\eta\,q^{1\over8}\,\widetilde{\chi}^\NS(z)\,\}
+\hf\,\{\chi_\hf^\NS(z)-h_3\,\eta\,q^{1\over8}\,\widetilde{\chi}^\NS(z)\}\nonumber\\
&=&\chi_\hf^\NS(z)\underbrace{-\hf \,(h_3+h_4)\,\eta\, q^{1\over8}}_{=g^{{\rm tw}}(\tau)}\,\widetilde{\chi}^{\NS}(z).
\ea

\subsection{The functions $f(\tau, \nu), 
g^{\rm inv}(\tau, \nu)$ and $g^{\rm tw}(\tau, \nu)$} \label{subappendix:BCDhnu}
From \eqref{Uhalf}--\eqref{Tzero} and using
\eqref{Uhalftheta}--\eqref{Tzerotheta} together with
\eqref{charactersh13}--\eqref{charactersh42} as well as some theta function 
identities obtained from \eqref{jacobiidentities2} and
\eqref{strange},
\begin{subequations}
\begin{multline}\label{Bnu}
f(\nu)
=-1+q^{1\over8}\frac{\vartheta_2^2\vartheta_3^2}{4\eta^3}\,\left\{\frac{\vartheta_4(\nu)^2}{\vartheta_1(\nu)^2}+\frac{\vartheta_3(\nu)^2}{\vartheta_2(\nu)^2}\right\}
-2q^{1\over8}\left\{\frac{\vartheta_2^4}{4\eta^3}- h_3 \eta\right\}\\
-(w+w^{-1})q^{1\over8}\,\frac{\vartheta_2^2\vartheta_3^2}{8\eta^3}\,\left\{\frac{\vartheta_4(\nu)^2}{\vartheta_1(\nu)^2}-\frac{\vartheta_3(\nu)^2}{\vartheta_2(\nu)^2}\right\},
\end{multline}
\begin{multline}
 g^{\rm inv}(\nu)
=q^{1\over8}\frac{\vartheta_2^2\vartheta_3^2}{2\eta^3}\,\left\{\frac{\vartheta_4(\nu)^2}{\vartheta_1(\nu)^2}-\frac{\vartheta_3(\nu)^2}{\vartheta_2(\nu)^2}\right\}
+(w+w^{-1})q^{1\over8}\,\left\{\frac{\vartheta_2^4}{2\eta^3}-2\eta h_3\right\}\\
-(w+w^{-1})q^{1\over8}\,\frac{\vartheta_2^2\vartheta_3^2}{4\eta^3}\,\left\{\frac{\vartheta_4(\nu)^2}{\vartheta_1(\nu)^2}
+\frac{\vartheta_3(\nu)^2}{\vartheta_2(\nu)^2}\right\},
\end{multline}
\be
g^{\rm tw}(\nu)
=q^{1\over8}\left\{ \frac{\vartheta_2^4}{4\eta^3}-h_3\eta\right\}-q^{1\over8}\frac{\vartheta_2^2\vartheta_3^2}{8\eta^3}\left\{ \frac{\vartheta_1(\nu)^2}{\vartheta_4(\nu)^2}+\frac{\vartheta_2(\nu)^2}{\vartheta_3(\nu)^2}\right\}.
\ee
\end{subequations}
 One now uses some of the relations \eqref{identityresidue} 
and \eqref{identityresidue2} between Appell functions as well as the theta function 
identities \eqref{jacobiidentities} and \eqref{jacobiidentities2} to obtain \eqref{Uhffinal}--\eqref{T0final}.

We also note that the functions $f(\nu)$, $g^{{\rm inv}}(\nu)$ and $g^{{\rm tw}}(\nu)$ 
may be written in terms of the massless characters of an $N=4$ SCA at central charge $c=6$. 
Taking advantage of  \eqref{magic}, \eqref{flowtotwist},
\eqref{sf} and \eqref{charN4h2h4h1z}, 
one has
\ba
f(\nu)
&=&-\hf \,\left[\frac{\chi_{1\over2}^{\Rat}(\nu)}{\widetilde{\chi}^{\Rat}(\nu)}
+\frac{\chi_{1\over2}^{\Ra}(\nu)}{\widetilde{\chi}^{\phantom{\widetilde{I}}\!\!\Ra}(\nu)}\right] 
-\hf (w+w^{-1})\left[\frac{\chi_0^{\Rat}(\nu)}{\widetilde{\chi}^{\Rat}(\nu)}
-\frac{\chi_0^{\Ra}(\nu)}{\widetilde{\chi}^{\phantom{\widetilde{I}}\!\!\Ra}(\nu)}\right],\\
g^{{\rm inv}}(\nu)
&=&- \left[\frac{\chi_{1\over2}^{\Rat}(\nu)}{\widetilde{\chi}^{\Rat}(\nu)}
-\frac{\chi_{1\over2}^{\Ra}(\nu)}{\widetilde{\chi}^{\phantom{\widetilde{I}}\!\!\Ra}(\nu)}\right] 
- (w+w^{-1})\left[\frac{\chi_0^{\Rat}(\nu)}{\widetilde{\chi}^{\Rat}(\nu)}+\frac{\chi_0^{\Ra}(\nu)}{\widetilde{\chi}^{\phantom{\widetilde{I}}\!\!\Ra}(\nu)}\right],\\
g^{{\rm tw}}(\nu)
&=&-\hf\,\left[\,\frac{\chi_{1\over2}^{\widetilde{\NS}}(\nu)}{\widetilde{\chi}^{\,\widetilde{\NS}}(\nu)}
+\frac{\chi_{1\over2}^{\NS}(\nu)}{\widetilde{\chi}^{\phantom{\widetilde{I}}\!\!\NS}(\nu)}\,\right].
\ea

\subsection{Fourier coefficients of 
$f(\tau, \nu)$ and $g^{{\rm tw}}(\tau, \nu)$} \label{subappendix:datafourier}
The data in 
Table \ref{tab:datafouriercoef} is presented in support of our claim that 
under deformation of our $\Zt$-orbifold CFT on K3 by $T^{\rm diag}$,
only $\qt$-BPS ground states stemming from the diagonal twisted sector
pair up with those in the untwisted subsector 
accounted for by
$U_{\ell=\hf}$ to form long representations off the BPS bound. 
For reference, we also present data
in Table \ref{tab:datafourierUzero} relating to the action of $\geo$ on $\qt$-BPS ground 
states from the untwisted subsector accounted for by $U_{\ell=0}$. 
We also note that for the action of the diagonal $SU(2)$
in $\geo\times\lgeo$
(see Section \ref{section:3} and Appendix \ref{subappendix:untwistedstates}), the information 
in Table \ref{tab:datafourierUzeroalt} should be used. However, at level 1 (resp. 2), the two 
triplets (resp. the two singlets)  from the untwisted subsector accounted for by $U_{\ell=\hf}$ 
 only  match the $SU(2)$ representation
contributions of the $\qt$-BPS ground states in the diagonal 
twisted sector: at level $n=1$, 
there are just no states available in the untwisted subsector $U_{\ell=0}$, and 
there are no singlets in that sector at level $n=2$.

\begin{sidewaystable}
\small{
\begin{tabular}{|l|llllllllllllllll|l|}
\hline 
&&&&&&&&&&&&&&&&&\\[-12pt]
\diagbox{$n$\phantom{g}}{$p$}&15&14&13&12&11&10&9&8&7&6&5&4&3&2&1&0&\\
\hline
&&&&&&&&&&&&&&&&&\\[-3pt]
1\,$(g^{{\rm tw}})$&&&&&&&&&&&&&&&$3^2$&&$g^{{\rm tw}}_1=6$\\
1\,$(f)$&&&&&&&&&&&&&&&$3^{\phantom{2}}$&&$f_1=3$\\[5pt]
2\,$(g^{{\rm tw}})$&&&&&&&&&&&&&&$5^{4}$&$3^{2}$&$1^{2}$&$g^{{\rm tw}}_{2}=28$\\
2\,$(f)$&&&&&&&&&&&&&&&&$1$&$f_{2}=1$\\[5pt]
3\,$(g^{{\rm tw}})$&&&&&&&&&&&&&$7^{6}$&$5^{6}$&$3^{8}$&$1^{2}$&$g^{{\rm tw}}_{3}=98$\\
3\,$(f)$&&&&&&&&&&&&&&$5^{3}$&$3$&&$f_{3}=18$\\[5pt]
4\,$(g^{{\rm tw}})$&&&&&&&&&&&&$9^{8}$&$7^{10}$&$5^{18}$&$3^{14}$&$1^{8}$&$g^{{\rm tw}}_{4}=282$\\
4\,$(f)$&&&&&&&&&&&&&&$5^{\phantom{3}}$&$3^{3}$&$1^{}$&$f_{4}=15$\\[5pt]
5\,$(g^{{\rm tw}})$&&&&&&&&&&&$11^{10}$&$9^{14}$&$7^{30}$&$5^{34}$&$3^{34}$&$1^{10}$&$g^{{\rm tw}}_{5}=728$\\
5\,$(f)$&&&&&&&&&&&&&$7^{5}$&$5^{3}$&$3^{5}$&$1^{3}$&$f_{5}=68$\\[5pt]
6\,$(g^{{\rm tw}})$&&&&&&&&&&$13^{12}$&$11^{18}$&$9^{42}$&$7^{60}$&$5^{76}$&$3^{58}$&$1^{28}$&$g^{{\rm tw}}_{6}=1734$\\
6\,$(f)$&&&&&&&&&&&&&$7^{3}$&$5^{9}$&$3^{7}$&$1^{2}$&$f_{6}=89$\\[5pt]
7\,$(g^{{\rm tw}})$&&&&&&&&&$15^{14}$&$13^{22}$&$11^{54}$&$9^{86}$&$7^{130}$&$5^{138}$&$3^{120}$&$1^{40}$&$g^{{\rm tw}}_{7}=3864$\\
7\,$(f)$&&&&&&&&&&&&$9^{7}$&$7^{9}$&$5^{15}$&$3^{14}$&$1^{6}$&$f_{7}=249$\\[5pt]
8\,$(g^{{\rm tw}})$&&&&&&&&$17^{16}$&$15^{26}$&$13^{66}$&$11^{112}$&$9^{188}$&$7^{240}$&$5^{272}$&$3^{204}$&$1^{86}$&$g^{{\rm tw}}_{8}=8182$\\
8\,$(f)$&&&&&&&&&&&&$9^{5}$&$7^{18}$&$5^{22}$&$3^{22}$&$1^{11}$&$f_{8}=358$\\[5pt]
9\,$(g^{{\rm tw}})$&&&&&&&$19^{18}$&$17^{30}$&$15^{78}$&$13^{138}$&$11^{246}$&$9^{354}$&$7^{468}$&$5^{472}$&$3^{380}$&$1^{134}$&$g^{{\rm tw}}_{9}=16618$\\
9\,$(f)$&&&&&&&&&&&$11^{9}$&$9^{15}$&$7^{30}$&$5^{45}$&$3^{39}$&$1^{13}$&$f_{9}=799$\\[5pt]
10\,$(g^{{\rm tw}})$&&&&&&$21^{20}$&$19^{34}$&$17^{90}$&$15^{164}$&$13^{304}$&$11^{468}$&$9^{688}$&$7^{824}$&$5^{856}$&$3^{634}$&$1^{252}$&$g^{{\rm tw}}_{10}=32550$\\
10\,$(f)$&&&&&&&&&&&$11^{7}$&$9^{30}$&$7^{50}$&$5^{67}$&$3^{60}$&$1^{24}$&$f_{10}=1236$\\[5pt]
11\,$(g^{{\rm tw}})$&&&&&$23^{22}$&$21^{38}$&$19^{102}$&$17^{190}$&$15^{362}$&$13^{582}$&$11^{914}$&$9^{1222}$&$7^{1482}$&$5^{1446}$&$3^{1104}$&$1^{392}$&$g^{{\rm tw}}_{11}=61828$\\
11\,$(f)$&&&&&&&&&&$13^{11}$&$11^{21}$&$9^{54}$&$7^{95}$&$5^{112}$&$3^{97}$&$1^{43}$&$f_{11}=2419$\\[5pt]
12\,$g^{{\rm tw}})$&&&&$25^{24}$&$23^{42}$&$21^{114}$&$19^{216}$&$17^{420}$&$15^{696}$&$13^{1140}$&$11^{1638}$&$9^{2198}$&$7^{2518}$&$5^{2474}$&$3^{1800}$&$1^{692}$&$g^{{\rm tw}}_{12}=114352$\\
12\,$(f)$&&&&&&&&&&$13^{9}$&$11^{42}$&$9^{87}$&$7^{144}$&$5^{181}$&$3^{150}$&$1^{55}$&$f_{12}=3780$\\[5pt]
13\,$(g^{{\rm tw}})$&&&$27^{26}$&$25^{46}$&$23^{126}$&$21^{242}$&$19^{478}$&$17^{810}$&$15^{1366}$&$13^{2054}$&$11^{2952}$&$9^{3756}$&$7^{4290}$&$5^{4054}$&$3^{3000}$&$1^{1074}$&$g^{{\rm tw}}_{13}=206526$\\
13\,$(f)$&&&&&&&&&$15^{13}$&$13^{27}$&$11^{78}$&$9^{160}$&$7^{245}$&$5^{286}$&$3^{240}$&$1^{92}$&$f_{13}=6801$\\[5pt]
14\,$(g^{{\rm tw}})$&&$29^{28}$&$27^{50}$&$25^{138}$&$23^{268}$&$21^{536}$&$19^{924}$&$17^{1592}$&$15^{2470}$&$13^{3714}$&$11^{5070}$&$9^{6402}$&$7^{7050}$&$5^{6678}$&$3^{4778}$&$1^{1786}$&$g^{{\rm tw}}_{14}=365232$\\
14\,$(f)$&&&&&&&&&$15^{11}$&$13^{54}$&$11^{129}$&$9^{252}$&$7^{386}$&$5^{436}$&$3^{359}$&$1^{146}$&$f_{14}=10659$\\[5pt]
15\,($g^{{\rm tw}}$)&$31^{30}$&$29^{54}$&$27^{150}$&$25^{294}$&$23^{594}$&$21^{1038}$&$19^{1818}$&$17^{2886}$&$15^{4476}$&$13^{6408}$&$11^{8662}$&$9^{10558}$&$7^{11572}$&$5^{10656}$&$3^{7690}$&$1^{2758}$&$g^{{\rm tw}}_{15}=633820$\\
15\,$(f)$&&&&&&&&$17^{15}$&$15^{33}$&$13^{102}$&$11^{240}$&$9^{426}$&$7^{608}$&$5^{694}$&$3^{551}$&$1^{208}$&$f_{15}=18137$\\[3pt]
\hline
\end{tabular}}
\caption{ $\geo$ representations and multiplicities at level $n, 1 \le n \le 15$. 
An entry of type $b^m$ at level $n$ is understood as multiplicity $m=g^{{\rm tw}}_{n,p}$ 
or $m=f_{n,p}$ of $\geo$ representation of dimension $b=2p+1$.}
\label{tab:datafouriercoef}
\end{sidewaystable}

\begin{table}
\begin{center}
\small{
\begin{tabular}{| l| l| l| l| l| l| l| l| l| l|}
\hline 
&&&&&&&&&\\[-12pt]
\diagbox{$n$\phantom{g}}{$p$}&15/2&13/2&1{1/2}&9/2&7/2&5/2&3/2&{1/2}&\\
\hline
&&&&&&&&&\\[-3pt]
1&&&&&&&$$&&$g^{{\rm inv}}_1=0$\\[3pt]
2&&&&&&&$4^4$&&$g^{{\rm inv}}_2=16$\\[3pt]
3&&&&&&&&$2^4$&$g^{{\rm inv}}_3=8$\\[3pt]
4&&&&&&$6^{8}$&$4^{4}$&$2^{4}$&$g^{{\rm inv}}_{4}=72$\\[3pt]
5&&&&&&$6^{4}$&$4^{12}$&$2^{4}$&$g^{{\rm inv}}_{5}=80$\\[3pt]
6&&&&&$8^{12}$&$6^{12}$&$4^{16}$&$2^{16}$&$g^{{\rm inv}}_{6}=264$\\[3pt]
7&&&&&$8^{8}$&$6^{24}$&$4^{28}$&$2^{20}$&$g^{{\rm inv}}_{7}=360$\\[3pt]
8&&&&$10^{16}$&$8^{24}$&$6^{44}$&$4^{56}$&$2^{32}$&$g^{{\rm inv}}_{8}=904$\\[3pt]
9&&&&$10^{12}$&$8^{48}$&$6^{72}$&$4^{76}$&$2^{60}$&$g^{{\rm inv}}_{9}=1360$\\[3pt]
10&&&$12^{20}$&$10^{36}$&$8^{84}$&$6^{132}$&$4^{140}$&$2^{92}$&$g^{{\rm inv}}_{10}=2808$\\[3pt]
11&&&$12^{16}$&$10^{72}$&$8^{132}$&$6^{204}$&$4^{224}$&$2^{136}$&$g^{{\rm inv}}_{11}=4360$\\[3pt]
12&&$14^{24}$&$12^{48}$&$10^{132}$&$8^{252}$&$6^{344}$&$4^{348}$&$2^{236}$&$g^{{\rm inv}}_{12}=8176$\\[3pt]
13&&$14^{20}$&$12^{96}$&$10^{216}$&$8^{392}$&$6^{540}$&$4^{540}$&$2^{344}$&$g^{{\rm inv}}_{13}=12816$\\[3pt]
14&$16^{28}$&$14^{60}$&$12^{180}$&$10^{400}$&$8^{656}$&$6^{864}$&$4^{860}$&$2^{524}$&$g^{{\rm inv}}_{14}=22368$\\[3pt]
15&$16^{24}$&$14^{120}$&$12^{300}$&$10^{624}$&$8^{1044}$&$6^{1320}$&$4^{1272}$&$2^{812}$&$g^{{\rm inv}}_{15}=34888$\\[3pt]
\hline
\end{tabular}}
\caption{$\geo$ representations and multiplicities at level $n, 1 \le n \le 15$. An entry of type $b^m$ at level $n$ is understood as multiplicity $m=g^{{\rm inv}}_{n,p}$ of $\geo$ representation of dimension $b=2p+1$.}
\label{tab:datafourierUzero}
\end{center}
\end{table}

\begin{table}
\begin{center}
\small{
\begin{tabular}{| l| l| l| l| l| l| l| l| l| l| l|}
\hline 
&&&&&&&&&&\\[-12pt]
\diagbox{$n$\phantom{g}}{$p$}&8&7&6&5&4&3&2&1&0&\\
\hline
&&&&&&&&&&\\[-3pt]
1&&&&&&&&&&$g^{{\rm inv}}_1=0$\\[3pt]
2&&&&&&&$5^{2}$&$3^{2}$&&$g^{{\rm inv}}_2=16$\\[3pt]
3&&&&&&&&$3^{2}$&$1^{2}$&$g^{{\rm inv}}_3=8$\\[3pt]
4&&&&&&$7^{4}$&$5^{6}$&$3^{4}$&$1^{2}$&$g^{{\rm inv}}_{4}=72$\\[3pt]
5&&&&&&$7^{2}$&$5^{8}$&$3^{8}$&$1^{2}$&$g^{{\rm inv}}_{5}=80$\\[3pt]
6&&&&&$9^{6}$&$7^{12}$&$5^{14}$&$3^{16}$&$1^{8}$&$g^{{\rm inv}}_{6}=264$\\[3pt]
7&&&&&$9^{4}$&$7^{16}$&$5^{26}$&$3^{24}$&$1^{10}$&$g^{{\rm inv}}_{7}=360$\\[3pt]
8&&&&$11^{8}$&$9^{20}$&$7^{34}$&$5^{50}$&$3^{44}$&$1^{16}$&$g^{{\rm inv}}_{8}=904$\\[3pt]
9&&&&$11^{6}$&$9^{30}$&$7^{60}$&$5^{74}$&$3^{68}$&$1^{30}$&$g^{{\rm inv}}_{9}=1360$\\[3pt]
10&&&$13^{10}$&$11^{28}$&$9^{60}$&$7^{108}$&$5^{136}$&$3^{116}$&$1^{46}$&$g^{{\rm inv}}_{10}=2808$\\[3pt]
11&&&$13^{8}$&$11^{44}$&$9^{102}$&$7^{168}$&$5^{214}$&$3^{180}$&$1^{68}$&$g^{{\rm inv}}_{11}=4360$\\[3pt]
12&&$15^{12}$&$13^{36}$&$11^{90}$&$9^{192}$&$7^{298}$&$5^{346}$&$3^{292}$&$1^{118}$&$g^{{\rm inv}}_{12}=8176$\\[3pt]
13&&$15^{10}$&$13^{58}$&$11^{156}$&$9^{304}$&$7^{466}$&$5^{540}$&$3^{442}$&$1^{172}$&$g^{{\rm inv}}_{13}=12816$\\[3pt]
14&$17^{14}$&$15^{44}$&$13^{120}$&$11^{290}$&$9^{528}$&$7^{760}$&$5^{862}$&$3^{692}$&$1^{262}$&$g^{{\rm inv}}_{14}=22368$\\[3pt]
15&$17^{12}$&$15^{72}$&$13^{210}$&$11^{462}$&$9^{834}$&$7^{1182}$&$5^{1296}$&$3^{1042}$&$1^{406}$&$g^{{\rm inv}}_{15}=34888$\\[3pt]
\hline
\end{tabular}}
\caption{$\geo\times\lgeo$ representations and multiplicities at level $n, 1 \le n \le 15$. 
An entry of type $b^m$ at level $n$ is understood as multiplicity $m=g^{{\rm inv}}_{n,p}$ 
of  the representation of dimension $b=2p+1$ of the diagonal
$SU(2)$ in $\geo\times\lgeo$.}
\label{tab:datafourierUzeroalt}
\end{center}
\end{table}
 \newpage
\section{Appendix - Quarter BPS states at level 2} \label{appendix:states}
\subsection{Twisted sector} \label{subappendix:twistedstates}
Note that $|\alpha_{\rm diag}\rangle$ is invariant under $\geo\times\lgeo$.
There  is a 28-dimensional space of massive $\qt$-BPS states in the diagonal twisted 
sector at level 2 which accounts for 
$g^{\rm tw}_2$ and which transforms trivially under $\lgeo$. 
We have already presented two $\geo$-singlets in 
\eqref{singletstwistedD2} which actually are the only singlets 
within   the 28-dimensional space in question. 
The remaining 26-dimensional space  transforms as  two triplets and four quintuplets
of $\geo$. We use the notations introduced in \eqref{Anotation} to give an explicit expression for these states.\\

\underline{\textbf{The two triplets:}}
\be
\{\, |t_1\rangle,\,|t_2\rangle,\,|t_3\rangle\,\}\qquad{\rm and}\qquad \{\, |\tilde{t}_1\rangle,\,|\tilde{t}_2\rangle,\,|\tilde{t}_3\rangle\,\},
\ee
with
\ba
 |t_1\rangle&=&\left(\aop_{-\frac{3}{2}}\,\ktm_0+3\aop_{-\frac{1}{2}}\,\ktm_{-1}+3\aop_{-\frac{1}{2}}\,\kop_{-1}\kom_0\ktm_0-6A_{++-}^{1112}-6A_{++-}^{1222}\right)|\alpha_{{\rm diag}}\rangle,\nonumber\\
|t_2\rangle&=&\left(\,\sum_{k=1}^2(-1)^k\left(\,(a_+^k)_{-\frac{3}{2}}(\chi_-^k)_0+3(a_+^k)_{-\frac{1}{2}}(\chi_-^k)_{-1}\right)+3\sum_{k=1}^2(a_+^k)_{-\frac{1}{2}}(\chi_+^{\widecheck{ k}})_{-1}(\chi_-^1)_{0}(\chi_-^2)_{0}\right.\nonumber\\
&&\phantom{spacefinder}\left.-6\sum_{k=1}^2(-1)^k\,(A_{++-}^{kkkk}+A_{++-}^{k\widecheck{k}\widecheck{k}k})\right)|\alpha_{{\rm diag}}\rangle,\nonumber\\
|t_3\rangle&=&\left(-\atp_{-\frac{3}{2}}\,\kom_0-3\atp_{-\frac{1}{2}}\,\kom_{-1}+3\atp_{-\frac{1}{2}}\,\ktp_{-1}\kom_0\ktm_0+6A_{++-}^{2221}+6A_{++-}^{2111}\right)|\alpha_{{\rm diag}}\rangle,\nonumber\\
\ea
and
\ba
|\tilde{t}_1\rangle&=&\left(\atm_{-\frac{3}{2}}\,\ktm_0+3\atm_{-\frac{1}{2}}\,\ktm_{-1}+3\atm_{-\frac{1}{2}}\,\kop_{-1}\kom_0\ktm_0-6A_{+--}^{2222}-6A_{+--}^{1122}\right)|\alpha_{{\rm diag}}\rangle,\nonumber\\
|\tilde{t}_2\rangle&=&\left(\,\sum_{k=1}^2\left(\,(a_-^k)_{-\frac{3}{2}}(\chi_-^{\widecheck{k}})_0+3(a_-^k)_{-\frac{1}{2}}(\chi_-^{\widecheck{k}})_{-1}\right)-3\sum_{k=1}^2\,(-1)^k\,(a_-^k)_{-
\frac{1}{2}}(\chi_+^k)_{-1}(\chi_-^1)_{0}(\chi_-^2)_{0}\right.\nonumber\\
&&\phantom{spacefinder}\left.-6\sum_{k=1}^2\,(A_{+--}^{kkk\widecheck{k}}+A_{+--}^{kk\widecheck{k}k})\right)|\alpha_{{\rm diag}}\rangle,\nonumber\\
|\tilde{t}_3\rangle&=&\left(\aom_{-\frac{3}{2}}\,\kom_0+3\aom_{-\frac{1}{2}}\,\kom_{-1}-3\aom_{-\frac{1}{2}}\,\ktp_{-1}\kom_0\ktm_0-6A_{+--}^{1111}-6A_{+--}^{2211}\right)|\alpha_{{\rm diag}}\rangle,\nonumber\\
\ea
where $\widecheck{k}:= 3-k$.\\

\underline{\textbf{The four quintuplets:}}
\ba
&&\left\{ A_{+++}^{1112}|\alpha_{{\rm diag}}\rangle,\,(A_{+++}^{1111}-3A_{+++}^{1122})|\alpha_{{\rm diag}}\rangle,\,(A_{+++}^{1222}-A_{+++}^{2111})|\alpha_{{\rm diag}}\rangle,\right.\nonumber\\[8pt]
&&\left.\phantom{spacechaserchaserchaserchas}(A_{+++}^{2222}-3A_{+++}^{2211})|\alpha_{{\rm diag}}\rangle,\,A_{+++}^{2221}|\alpha_{{\rm diag}}\rangle\,\right\},\nonumber\\[10pt]
&&\left\{ A_{---}^{2222}|\alpha_{{\rm diag}}\rangle,\,(A_{---}^{2221}+3A_{---}^{1222})|\alpha_{{\rm diag}}\rangle,\,(A_{---}^{1122}+A_{---}^{2211})|\alpha_{{\rm diag}}\rangle,\right.\nonumber\\[8pt]
&&\left.\phantom{spacechaserchaserchaserchas}(A_{---}^{1112}+3A_{---}^{2111})|\alpha_{{\rm diag}}\rangle,\,A_{---}^{1111}|\alpha_{{\rm diag}}\rangle\,\right\},\nonumber\\[10pt]
&&\left\{ A_{++-}^{1122}|\alpha_{{\rm diag}}\rangle,\,(A_{++-}^{1112}+A_{++-}^{1121}-2A_{++-}^{1222})|\alpha_{{\rm diag}}\rangle,\,(A_{++-}^{1111}+A_{++-}^{2222}-2A_{++-}^{2112}-2A_{++-}^{1221})|\alpha_{{\rm diag}}\rangle,\right.\nonumber\\[8pt]
&&\left.\phantom{spacechaserchaserchaserchas}(A_{++-}^{2221}+A_{++-}^{2212}-2A_{++-}^{2111})|\alpha_{{\rm diag}}\rangle,\,A_{++-}^{2211}|\alpha_{{\rm diag}}\rangle\,\right\},\nonumber\\[10pt]
&&\left\{ A_{+--}^{1222}|\alpha_{{\rm diag}}\rangle,\,(A_{+--}^{2222}-A_{+--}^{1221}-2A_{+--}^{1122})|\alpha_{{\rm diag}}\rangle,\,(A_{+--}^{1112}-A_{+--}^{2221}+2A_{+--}^{1211}-2A_{+--}^{2122})|\alpha_{{\rm diag}}\rangle,\right.\nonumber\\[8pt]
&&\left.\phantom{spacechaserchaserchaserchas}(A_{+--}^{1111}-A_{+--}^{2112}-2A_{+--}^{2211})|\alpha_{{\rm diag}}\rangle,\,A_{+--}^{2111}|\alpha_{{\rm diag}}\rangle\,\right\}.
\ea

\subsection{Untwisted sector} \label{subappendix:untwistedstates}
Besides the two singlets \eqref{singletsuntwistedB2}, the untwisted sector at level 2 contains a
$16$-dimensional space of massive $\qt$-BPS states contributing to $g^{\rm inv}_2$. 
They are $N=4$ primaries built as eight odd combinations of oscillator modes acting on the 
$\Zt$-orbifold odd ground states 
$\widetilde{\Omega}^1:=\kbop_{-\hf}\,\Omega$ and 
$\widetilde{\Omega}^2:=\kbtp_{-\hf}\,\Omega$. 
These two states transform non-trivially under our geometric symmetry groups $G_i$.
Indeed, as argued in Section \ref{section:3}, 
they are invariant under the action of $\geo$ and form a doublet under the action 
of $\lgeo$.

On the other hand,  we  find four quadruplet representations of $\geo$ at level $2$ in this
sector. We introduce the notation
\be
B_\pm^{k\ell m}:=(a_\pm^k)_{-1}\,(\chi_+^\ell)_{-\hf}\,(\chi_-^m)_{-\hf},\qquad k, \ell, m \in \{1,2\},
\ee
so that the $16$-dimensional space is generated by the following  $16$ states:
\be
\begin{array}{l}
B_+^{112}\widetilde{\Omega}^i,\;\;
(B_+^{122}+B_+^{212}-B_+^{111})\widetilde{\Omega}^i,\;\;
(B_+^{211}+B_+^{121}-B_+^{222})\widetilde{\Omega}^i,\;\;
\,B_+^{221}\widetilde{\Omega}^i,\\[5pt]
B_-^{212}\widetilde{\Omega}^i,\;\;
(B_-^{112}+B_-^{211}-B_-^{222})\widetilde{\Omega}^i,\;\;
(B_-^{221}+B_-^{122}-B_-^{111})\widetilde{\Omega}^i,\;\;
B_-^{121}\widetilde{\Omega}^i.
\end{array}
i \in \{1,2\},
\ee
where each row generates a $(4,2)$ representation of 
$\geo\times\lgeo$.
Note that under the action of the diagonal subgroup $SU(2)$  of
$\geo\times\lgeo$, we obtain two copies of 
$\textbf{4}\otimes\textbf{2}\cong\textbf{3}\oplus\textbf{5}$, yielding no 
singlets altogether.


\acknowledgments
We gratefully acknowledge support from the Simons Center for Geometry and Physics, 
Stony Brook University, at which a major part of the research for this paper was performed.
AT thanks the Mathematics Institute at the University of Freiburg for their warm welcome and the
support received from the DFG Research training  group GRK 1821 ``Cohomological methods in geometry''.
We also thank Christoph Keller and Ida Zadeh 
for helpful correspondences. 
An anonymous referee deserves our thanks for diligent reading and 
helpful suggestions.

\end{document}